\documentclass[12pt]{article}  
\usepackage{epsfig}
\usepackage{mathbbol}

\begin{document}

\renewcommand{\baselinestretch}{1.0}
\textwidth 149mm
\textheight 220mm
\topmargin 0pt
\oddsidemargin 5mm

\def\thefootnote{\fnsymbol{footnote}}

\newcommand{\eq}{\begin{equation}}
\newcommand{\en}{\end{equation}}
\newcommand{\eqa}{\begin{eqnarray}}
\newcommand{\ena}{\end{eqnarray}}
\newcommand{\spz}{\hspace{0.7cm}}
\newcommand{\lbl}{\label}
\newcommand{\lhi}{\hat\lambda_{i}}
\newcommand{\br}{\langle}
\newcommand{\kt}{\rangle}
\newcommand{\um}{\frac12}
\newcommand{\mc}{\multicolumn}

\begin{titlepage}
\vskip0.5cm
\begin{flushright}
SHEP-0413 \\
CPT-2004/P.024 \\
DESY 04-090\\
\end{flushright}
\vskip0.5cm
\begin{center}
{\Large\bf ${\rm SU(3)}$ lattice gauge theory with a mixed fundamental 
 and adjoint plaquette action: \\
Lattice artefacts \\
}
\end{center}
\vskip 0.3cm

\centerline{
M. Hasenbusch$^{a}$ and S. Necco$^{b}$
}
\vskip 0.4cm
\centerline{\sl $^{a}$
 Department of Physics and Astronomy, University of 
Southampton,} 
\centerline{\sl
Southampton SO17 1BJ, United Kingdom}
\vskip 0.3cm
\centerline{\sl $^{b}$ Centre de Physique Th\'eorique }
\centerline{\sl CNRS Luminy, Case 907 F-13288, Marseille Cedex 9, France}
\vskip 0.3cm
\centerline{e-mail: hasenbus@phys.soton.ac.uk, necco@cpt.univ-mrs.fr}

\vskip 0.4cm
\begin{abstract}
We study the four-dimensional ${\rm SU(3)}$ gauge model with a fundamental
and an adjoint plaquette term in the action.
We investigate whether corrections to scaling
can be reduced by using a negative value of the adjoint coupling. 
To this end, we  have studied
the finite temperature phase transition, the static
potential and the mass of the $0^{++}$ glueball. In order to compute 
these quantities we have implemented  variance reduced estimators 
that have been proposed recently.
Corrections to scaling 
are analysed in dimensionless combinations such as $T_c/\sqrt{\sigma}$ and
$m_{0^{++}}/T_c$.  We find that indeed the lattice artefacts
in e.g. $m_{0^{++}}/T_c$ can be reduced considerably compared with the 
pure Wilson (fundamental) gauge action at the same lattice spacing.
\end{abstract}
\end{titlepage}

\section{Introduction}
Due to the enormous effort required by lattice QCD simulations, 
one is restricted 
to rather coarse lattice spacings. Therefore it is important to choose
the lattice action such that already at a coarse lattice spacing the
lattice artefacts are small. In the past, various proposals had been made 
to this end.

One approach are so called perfect actions (for recent work on this
subject see e.g. ref. \cite{perfect}). Here one tries
to avoid lattice artefacts completely, at the price of, in principle, 
an infinite number of terms in the action. 
In order to make this
approach work practically, the set of terms has to be truncated
to a finite, still large number of terms. The error introduced
by this truncation is hard to estimate a priori.

Alternatively, Symanzik \cite{SYM} has proposed a scheme that allows
to eliminate lattice artefacts systematically, order by order
in the lattice spacing and the coupling constant. For the SU(3)
lattice gauge theory this has been worked out by L\"uscher and Weisz
\cite{LuWe85} up to the 1-loop level in perturbation theory to eliminate
$O(a^2)$ corrections.

Since we work at rather large values of the gauge-coupling it is not
clear a priori, whether the 1-loop improvement of ref. \cite{LuWe85} is of any
help. Only studies (see e.g. ref. \cite{SN03})
of the scaling behaviour of various quantities can clarify this question.

Here, we pursue a rather ad hoc  approach. 
It is well established that the pure SU(3) gauge theory in four dimensions 
has a line of first order phase transitions with an end-point at 
\cite{Blum:1995xb}
\begin{equation}
\label{endpoint}
 (\beta_f,\beta_a) = (4.00(7),2.06(8)) \;\;\;.
\end{equation}
For the precise definition of $\beta_f$ and $\beta_a$ see 
eq.~(\ref{mixed_action}) below.
At such transitions, 
lattice artefacts might completely disguise the 
continuum physics. Here, in particular, the mass 
$m_{0^{++}}$ of the lightest glueball $0^{++}$  goes to zero as the end-point,
eq.~(\ref{endpoint}), is approached \cite{Heller}. As a relic of this 
behaviour, 
the estimate of $m_{0^{++}}/T_c$ ($T_c$ is the location of the finite 
temperature phase transition)
at $\beta_a=0$, for $5.5 < \beta_f < 6.0$ is much 
smaller than the continuum result.
Here, we study whether the scaling behaviour can be improved 
by staying apart from the transition line
by choosing  negative values of $\beta_a$.
Recently one of us \cite{SN03} has studied 
gauge actions with $2 \times 1$ - Wilson loops in addition to 
the plaquette. These actions are motivated by 
the RG-group and the Symanzik improvement programme. Indeed, for the 
so called Iwasaki action \cite{Iwasaki} an improved scaling 
behaviour was observed.
Here we investigate whether
by adding the adjoint part to the action similar
benefits can be achieved, while keeping the action as local as possible;
i.e. using plaquette terms only.
For the mixed fundamental-adjoint action, 
it is straightforward  to construct a transfer matrix along 
the lines of ref. \cite{transfermatrix}. However, for our choices 
of $\beta_a$, it is not positive.  For a detailed discussion see the 
appendix.

An other important question (that we do not study here) is, 
whether dislocations, i.e. topological 
objects that are pure lattice artefacts, can be suppressed by a modification 
of the standard Wilson gauge action. 


The outline of our paper is as follows:
First we give the definition of the model and the observables that we 
have studied. Next we discuss the Monte Carlo algorithm that was used to 
generate the gauge-configurations. In section 4 we study the finite 
temperature phase transition. In section 5 we extract the string tension
$\sigma$ from the Polyakov loop correlation function. To this end we have 
implemented a variant of the variance reducing algorithm
proposed by L\"uscher and Weisz \cite{lw}.
Next, in section 6, 
we determine the mass $m_{0^{++}}$ of the lightest
glueball $0^{++}$. 
Also here, we have employed a new algorithm \cite{meyer}, to reduce 
the variance of correlation functions.
Based on these results, 
we compute the dimensionless quantities $T_c/\sqrt{\sigma}$ and 
$m_{0^{++}}/T_c$.

\section{The model}
We consider a hypercubical lattice with the linear extension $a N_s$ in 
the spacial directions and $a N_t$ in the temporal direction. $a$ is the 
lattice spacing. $N_s$ and $N_t$ are integer numbers. In all directions, 
periodic boundary conditions are applied. 

The action describing the pure SU(N) lattice gauge theory containing
plaquette terms only, can be written in the general form \cite{GoKo}
\begin{equation}
S \;=\;\sum_{\alpha} \tilde{\beta_{\alpha}} \; \sum_P 
\left[1-\frac{1}{d(\alpha)} \mbox{Re} \mbox{Tr}_{\alpha} 
U_P \right] \;\;,
\end{equation}
where the sum over $\alpha$ indicates the sum over all representations of
SU(N), $\tilde{\beta_{\alpha}}$ is the coupling associated 
with each representation,
$d(\alpha)$ is the dimension of the representation, 
$\mbox{Tr}_{\alpha}$ is the trace in the given representation and
$U_P$ is the path ordered product of the link variables along an elementary
plaquette $P$:
\begin{equation}
U_P(x,\mu,\nu) = U_{x,\mu} U_{x+\hat \mu,\nu} U_{x+\hat \nu,\mu}^{\dag} 
U_{x,\nu}^{\dag}\;\;,
\end{equation}
where $x$ labels the sites of the lattice and $\hat \mu$ is a unit
vector in the $\mu$-direction.

In particular, we will consider the mixed fundamental-adjoint action
\begin{equation}
S \;=\; \tilde{\beta_f}\; \sum_P \left[1-\frac{1}{N} \mbox{Re} \mbox{Tr}_f U_P
\right]+
\tilde{\beta_a}\; \sum_P \left[1-\frac{1}{N^2-1} \mbox{Re} \mbox{Tr}_a U_P
\right] \;\;. 
\end{equation}
Using the identity
\begin{equation}
\mbox{Tr}_a U = \mbox{Tr}_f U^{\dag} \mbox{Tr}_f U \; -\; 1 
\end{equation}
and defining
\begin{equation}
\beta_f=\tilde{\beta_f},\quad \beta_a=\frac{N^2}{N^2-1}\tilde{\beta_a}\;\;,
\end{equation}
one obtains
\begin{equation}\label{mixed_action}
S \;=\; \beta_f \; \sum_P \left[1-\frac{1}{N} \mbox{Re} \mbox{Tr}_f U_P \right]
      \;+\; \beta_a \; \sum_P
 \left[1-\frac{1}{N^2} \mbox{Tr}_f U_P^{\dag} \mbox{Tr}_f U_P \right]\;\;,
\end{equation}
which is the form that we will use in the following for SU(3).
In the naive continuum limit the bare coupling $g_0$ for the 
action eq.~(\ref{mixed_action}) is given by
\begin{equation}
\label{constant_tree_level}
\frac{6}{g_0 ^2}=\beta_{W}=\beta_f+2\beta_a \;\;,
\end{equation}
where we have introduced, following the literature, the notion of an 
``equivalent Wilson coupling'' $\beta_W$.

%
\subsection{Observables} 
In the following we define the basic observables that we have studied.
The fundamental plaquette is given by
\begin{equation}
E_f =\frac{1}{6 V} 
\sum_{x,\mu,\nu>\mu} \langle \mbox{Tr}_f U_P
\rangle \;\;,
\end{equation}
where $V=N_{s}^3 N_{t}$ is the number of points of the  lattice.\\
The adjoint plaquette can be defined by
\begin{equation}
E_a = 
       \frac{1}{18 V} 
\sum_{x,\mu,\nu>\mu} \langle \mbox{Tr}_f  U^{\dagger}_P 
 \mbox{Tr}_f U_P  \rangle \;\;.
\end{equation}
To study the finite temperature phase transition and the potential 
of static quarks we consider the Polyakov loop
\begin{equation}
 P_{\vec{x}} = \mbox{Tr}_f \prod_t U_{\vec{x},t,0} \;\;,
\end{equation}
where $U_{\vec{x},t,\mu}$ is the link variable starting in $\vec{x},t$ in the
direction $\mu$. 
In particular, we study the histogram associated with the quantity
\begin{equation}
\label{omega}
 \Omega = \frac{1}{N_s^3} \sum_{\vec{x}}  P_{\vec{x}} \;\;.
\end{equation}

\section{Simulation algorithm}
We have simulated the model with a local Metropolis procedure. 
The proposal for a new value of a link variable 
is generated by a Cabbibo-Marinari (CM) heat-bath update 
\cite{cama}
of a single 
SU(2)-subgroup for the action
\begin{equation}
 S_0 \;=\; \beta_f' \; 
\sum_P \left[1-\frac{1}{N} \mbox{Re} \mbox{Tr}_f U_P \right] \;\;.
\end{equation}
This proposal is then accepted with the probability 
\begin{equation}
 A = \mbox{min}\left[1,\exp(-S(U')+S_0(U')+S(U)-S_0(U))\right] \;\;,
\end{equation}
where $U$ is the original gauge field and $U'$ is the proposal.
$\beta_f'<\beta_f$ 
is tuned such that the optimal acceptance rate is obtained.
In addition, we have performed overrelaxation (OV) updates \cite{over}
that keep the 
fundamental part of the action constant. Here, we accepted the proposal
with the probability
\begin{equation}
 A = \mbox{min}\left[1,\exp(-S_a(U')+S_a(U))\right] \;\;,
\end{equation}
where 
\begin{equation}
S_a(U)=\; \beta_a \; \sum_P
 \left[1- \frac{1}{N^2} \mbox{Tr}_f U_P^{\dag} \mbox{Tr}_f U_P \right]\;.
\end{equation}
In both cases we applied, for a given link, sub-group updates for 
the 1-2, 1-3 and 2-3 components in a sequence.

Using these elementary link-updates we are sweeping over the lattice.
A complete update cycle is given by 
\begin{itemize}
\item One Cabbibo-Marinari-Metropolis sweep
\item $M$ overrelaxation-Metropolis sweeps.
\end{itemize}

We have implemented the algorithm in C as well as in Fortran.
On the one hand, this is a good check for the correctness of the
implementation and on the other, for the determination of the glueball
mass, we intended to use part the Fortran code that was written for the 
study reported in ref. \cite{SN03}. 
We have used the random number generator discussed in ref. \cite{Lrandom}.
For both implementations,
we find that the time needed
to update a link variable for the mixed action is roughly twice the 
time needed for the Wilson gauge action (i.e. $\beta_a=0$). 
With the C-program, we need $1.4 \times 10^{-5} s$ (CM-Metropolis) 
and $0.9 \times 10^{-5} s$ (OV-Metropolis)
for the update of a single link variable  on a Pentium 4 PC 
running at $1.7$ GHz. 

\subsection{Comparison with the literature}
As a test of the correctness of the program we tried to reproduce the 
values for the fundamental and adjoint plaquette given in fig. 2 and 
Table 1 of ref. \cite{Blum:1995xb}. 

From fig. 2 one reads off that $E_f \approx 1.35$ for $\beta_f=3.8$,
$\beta_a=2.25$ and $E_f \approx 1.88$ for $\beta_f=4.0$, $\beta_a=2.25$.
Simulating an $8 \times 12^3$ lattice we find  $E_f=1.3465(12)$ and 
$E_f=1.8803(4)$, respectively. In both cases,  
700 update cycles each with one CM-Metropolis sweep followed by M=5 
OV-Metropolis sweeps were performed. We have discarded 200 and 100  cycles,
respectively.

In Table 1 of ref. \cite{Blum:1995xb}, the authors give values for the 
jump of the fundamental and adjoint plaquette at the first order 
bulk phase transition. Among other values, they give
$\Delta E_f = 0.656(2) $ and $\Delta E_a = 0.464(2)$ at
$\beta_f=3.27$, $\beta_a=3.0$.

We computed these values on a $8 \times 12^3$ lattice. 
We have started one simulation with an ordered 
and one with a disordered
configuration. 
Within the simulation we did not observe tunneling between 
the phases.
Hence we computed $\Delta E_f$ and $\Delta E_a$ as the difference 
of $E_f$ and $E_a$ obtained from the run with ordered and the run with 
disordered start. Our result is $\Delta E_f=1.8863(7)-1.2287(10)=0.6576(12)$
and $3 \Delta E_a=3.947(2)-2.550(2)=1.397(3)$ in perfect agreement with 
ref. \cite{Blum:1995xb}. Both simulations had 400 cycles with 50 cycles 
discarded for equilibration.

\subsection{Tuning the algorithm}
\begin{table}
\caption{\sl \label{accept}
Acceptance rate as a function of $\beta_f'$. We have simulated a
 $4 \times 16^3$ lattice at $\beta_f=9.25$ and $\beta_a=-4.0$. 
An update-cycle consists of one CM-Metropolis sweep followed by $5$ 
OV-Metropolis sweeps. In each case, we performed 
20000 cycles with 
1000 discarded for thermalisation.
}
\begin{center}
\begin{tabular}{|l|l|l|l|l|l|}
\hline
\multicolumn{1}{|c}{$\beta_f'$}&
\multicolumn{1}{|c}{$P_{acc,heat}$}  & 
\multicolumn{1}{|c}{$P_{acc,over}$} & 
\multicolumn{1}{|c}{$E_f$} & 
\multicolumn{1}{|c}{3 $E_a$} & 
\multicolumn{1}{|c|}{$|\Omega|$}\\
\hline
    4.0    &  0.76799(5)  &0.80855(4)   & 1.6146(2)&2.9472(5) &0.056(5)\\      
    5.0    &  0.83647(4)  &0.80861(4)   & 1.6148(2)&2.9478(5) &0.062(5)\\
    6.0    &  0.82884(1)  &0.80854(3)   & 1.6145(1)&2.9469(4) &0.055(4)\\ 
    7.0    &  0.77845(2)  &0.80856(3)   & 1.6145(1)&2.9471(4) &0.056(4) \\
    9.25   &  0.64654(2)  &0.80855(3)   & 1.6145(1)&2.9470(4) &0.054(4)\\
\hline
\end{tabular}
\end{center}
\end{table}

To this end, we have simulated  a $4 \times 16^3$ lattice at
$\beta_f=9.25$ and $\beta_a=-4.0$. In Table \ref{accept} we give the acceptance
rate of the OV-Metropolis and the CM-Metropolis step as a
function of $\beta_f'$.  We see that in the case of the CM-Metropolis
the acceptance rate depends very weakly on $\beta_f'$.
The optimal value is reached for $\beta_f'$ around $5.0$ to $6.0$. 
The acceptance rate for the OV-Metropolis is the same for all runs, as it
should. With more than $80\%$ it is still reasonable.


\section{The finite temperature phase transition}
\label{finiteTsection}
The pure Yang-Mills theory undergoes a first order phase transition at a finite
temperature $T_{c}$ \cite{Polyakov:1978vu,Susskind:1979up}.
In the high temperature phase, the system is disordered and in the 
thermodynamic limit, the expectation value of the Polyakov loop vanishes.
On the other hand, in the low temperature phase, the $Z_3$ centre 
symmetry is broken.\\
For a lattice with the extension of $N_t$ lattice spacings
in the time direction, in the 
limit $N_s \rightarrow \infty$,
the deconfinement temperature 
is given by
\begin{equation}
\frac{1}{T_c}=N_t a(\{\beta_f,\beta_a\}_c) \;\;,
\end{equation}
where $\{\beta_f,\beta_a\}_c$ indicates the critical coupling. In our
numerical study, we could not compute the complete critical curve
$\{\beta_f,\beta_a\}_c$ for a given value of $N_t$, but instead, we have 
only determined $\beta_{f,c}$ for the fixed values $\beta_a=0,-2$ and $-4$
of the adjoint coupling constant.


In order to determine $\beta_{f,c}$, 
we have studied the probability distribution $p(|\Omega|)$  
(for the definition of $\Omega$ see eq.~(\ref{omega})).  
At the transition point, for sufficiently large lattice sizes, 
the histogram of the order parameter has a double peak structure.
The weight of each of the phases should be the same at the
transition point (see ref. \cite{Borgs}).
One should note that, due to the $Z_3$ centre symmetry, 
the ordered phase is threefold degenerate. 
On the finite lattice we need some rule, how to assign given configurations
to a particular phase. Here, we assign configurations with 
$|\Omega| < O_{min}$ to the disordered phase, and  configurations with
$|\Omega| > O_{min}$ to the ordered phase, where $O_{min}$ is the 
minimum of $p(|\Omega|)$ between the two peaks. 
As the lattice size increases, the separation of the peaks becomes sharper 
and therefore, the ambiguity of this assignment vanishes.

In particular, we have 
computed the weight of the disordered phase as 
\begin{equation}
\label{dis}
 P_{dis} = \int_{0}^{O_{min}} \mbox{d} |\Omega| \;  p(|\Omega|)
\end{equation}
and the weight of the three ordered phases as
\begin{equation}
\label{order}
 P_{order} = \int_{O_{min}}^{\infty} \mbox{d} |\Omega| \; p(|\Omega|) \;\;.
\end{equation}

The estimate of the critical $\beta_{f,c}$ is given by
\begin{equation}
\label{defbetac}
\frac{P_{order}(\beta_{f,c})}{P_{dis}(\beta_{f,c})} = 3 \;\;,
\end{equation}
where the factor 3 stems from the threefold degeneracy of the 
ordered phase.
Following ref. \cite{Borgs},  the
estimates for $\beta_{f,c}$ should converge 
exponentially fast as $N_s$ is increasing.

In the Monte Carlo simulation,  probabilities such as  
eq.s~(\ref{dis},\ref{order}) can be obtained in the following way:
the probability that $|\Omega|$ is in the interval $[O_1,O_2]$ is estimated by
\begin{equation}
 P[O_1,O_2] \approx \frac{1}{N_{max}-N_{disc}} \sum_{i=N_{disc}+1}^{N_{max} } 
     \chi_{[O_1,O_2]}(|\Omega|^{(i)})     \;\;,    
\end{equation}
where $\chi_{[O_1,O_2]}$ is the characteristic function of the interval 
$[O_1,O_2]$. $|\Omega|^{(i)}$ is the value of $|\Omega|$ for the $i^{th}$
configuration, $N_{max}$ is the total number of configurations that
has been generated and $N_{disc}$ the number of configurations that 
have been discarded for equilibration. I.e., we just count how frequently
the observable falls into the desired interval.




Using reweighting we can easily obtain the histogram of $|\Omega|$  for
all values of $\beta_f$ in the neighborhood of $\beta_{f,s}$, where the 
simulation has been performed  (i.e. the configurations 
are generated with the Boltzmann weight for $\beta_{f,s}$). 
The standard reweighting method gives in our case:
\begin{equation}
\label{reweighting}
 P[O_1,O_2](\beta_f) \approx 
 \frac{\sum_{i=N_{disc}+1}^{N_{max}}
 \chi_{[O_1,O_2]}(|\Omega|^{(i)})
\exp(-[\beta_{f}-\beta_{f,s}] \sum_{x,\mu,\nu>\mu} \mbox{Re} \mbox{Tr}_f U_P^{(i)} ) }
{
\sum_{i=N_{disc}+1}^{N_{max}}
\exp(-[\beta_{f}-\beta_{f,s}] \sum_{x,\mu,\nu>\mu} \mbox{Re} \mbox{Tr}_f U_P^{(i)} ) }
\;\;.
\end{equation}
If $\beta_{f,s}$ is close enough 
to $\beta_{f,c}$, i.e. a double peak can be seen 
at $\beta_{f,s}$,  we use eq.~(\ref{reweighting})
in combination with the intersection method to find the solution of  
eq.~(\ref{defbetac}).

Mostly, we have started our search for $\beta_{f,c}$ on lattices 
with $N_s=3 N_t$. 
Here it is helpful, to have a good first guess for $\beta_{f,c}$ that 
can be used as a first simulation point $\beta_{f,s}$. To this end, 
see our discussion on lines of constant physics in subsection \ref{linesof}.

The result for $\beta_{f,c}$
obtained with $N_s=3 N_t$ was then used as a first guess $\beta_{f,s}$
for the simulation of the next larger lattice size and so on. 
For lattice sizes $N_s < 6 N_t$ we have used the Monte Carlo algorithm
as described in the previous section for the simulation. 
For $N_s = 6 N_t$, where our 
final results for $\beta_{f,c}$ are taken from, we have employed
the multicanonical method \cite{BeNe92} on top of it. A discussion 
is given in the following subsection.

\subsection{Enhancing the Tunneling Rate}
As the lattice size increases, the separation of the phases becomes 
more pronounced. While this allows for an unambiguous
separation of the phases and hence of the transition temperature, 
it has adverse effects on the Monte Carlo simulation itself. 
Since in each elementary step of the update, the configuration is only changed
by a little, i.e. just at a single link, going form an ordered 
phase to the disordered, we have to pass through configurations that
are mixtures of the phases. However, if such configurations 
are strongly suppressed, the Monte Carlo time to go from one phase 
to the other will be very large. In order to overcome this problem,
it has been proposed not to generate the configurations with their
Boltzmann-weight but rather with some modified one.
In order to enhance the rate of tunneling events from one phase to the 
other we have simulated with a multicanonical ensemble \cite{BeNe92}.
I.e. we have generated the configurations with a weight proportional to
\begin{equation}
 \tilde B[U] = \exp(-S[U]) \; W(|\Omega|) \;\;,
\end{equation}
where the modification factor $W(|\Omega|)$ only depends on the 
modulus of the sum over the Polyakov loops.

We sweep over the whole lattice with the CM-Metropolis and the 
OV-Metropolis as described above. In order to fulfil detailed balance, 
with a probability of $50\%$ the update sweep is performed in exactly 
reversed order. After these sweeps, an accept/reject step is performed 
with the acceptance 
\begin{equation}
\label{accmodified}
 A = \mbox{min}[1,W(|\Omega'|)/W(|\Omega|)| \:\:\;,
\end{equation}
where $|\Omega'|$ is the value of $|\Omega|$ for the proposed 
configuration.
For the modification factor $W(|\Omega|)$, we have used the ansatz
\begin{eqnarray}
W(|\Omega|) &=&  1  \phantom{xxxxxxxxxxxxxxxxxxxxx} \mbox{for} 
\;\;\;\;\;\;\; \phantom{s_0 \le} |\Omega| < s_1 \nonumber
\\
W(|\Omega|) &=& 1+m \; (|\Omega|-s_1)/(s_2-s_1) \;\;\;\;  \mbox{for} \;\;\;\;\;\; 
 s_1 \le |\Omega| < s_2 \nonumber
\\
W(|\Omega|) &=&  1+m \phantom{xxxxxxxxxxxxxxxxx} \;  \mbox{for} \;\;\;\;\;\; s_2 \le |\Omega| < s_3 \nonumber
\\
W(|\Omega|) &=&  1+m\; (s_4-|\Omega|)/(s_4-s_3) \;\;\;\;   \mbox{for} \;\;\;\;\;\;
s_3 \le |\Omega| < s_4 \nonumber
\\
W(|\Omega|) &=&  1  \phantom{xxxxxxxxxxxxxxxxxxxxx} \mbox{for} \;\;\;\;\;\; 
s_4 \le 
|\Omega| \;\;. 
\end{eqnarray}
The parameters $s_1,s_2,s_3,s_4,s_5$ and $m$ of this ansatz were essentially
determined by trail and error.  We have used this method only for our largest 
value of $N_s$: $N_s=6 N_t$ for a given temporal extension $N_t$. 

\subsection{Numerical results for $\beta_{f,c}$}
Let us discuss in detail the 
example of a $6 \times 36^3$ lattice simulated at 
$\beta_f=7.803$ and $\beta_a=-2.0$.  
For this simulation, we have chosen
the parameters of $W$ as  $s_1=0.03$, $s_2=0.06$, $s_3=0.08$, $s_4=0.15$, and
$m=4.0$.

\begin{figure}
 \begin{center}
 \includegraphics[width=11cm]{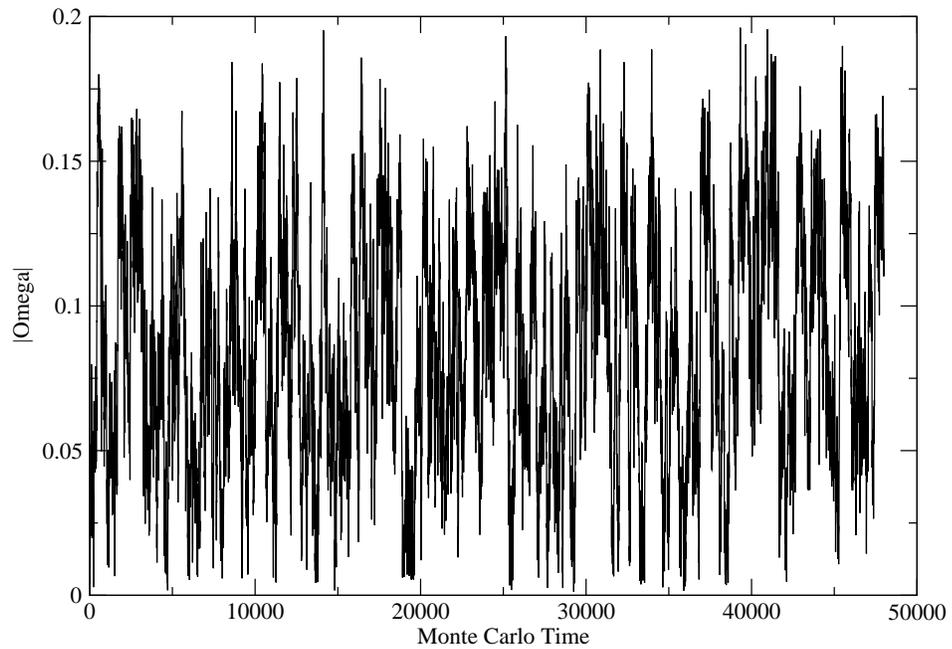}
 \end{center}
 \caption{Evolution of $|\Omega|$ in Monte Carlo time. A unit in time
 is given by a cycle, consisting of one CM-Metropolis and 5 OV-Metropolis
 sweeps. We have simulated a
 $6 \times 36^3$ lattice, at $\beta_f=7.803$, $\beta_a=-2.0$, with the 
 modified weight that is specified in the text.
 }\label{history}
\end{figure}

The simulation consists of 
48000 update-cycles with one CM-Metropolis and five OV-Metropolis sweeps
each. The acceptance rate for the modified weight eq.~(\ref{accmodified}) was 
about $93 \%$.  In fig. \ref{history} the evolution of $|\Omega|$ in 
Monte Carlo time is shown.  We see that the system indeed tunnels quite 
frequently from the disordered phase into one of the ordered phases.
For equilibration, we have discarded the first 2000 update-cycles.

\begin{figure}
\begin{center}
\includegraphics[width=10cm]{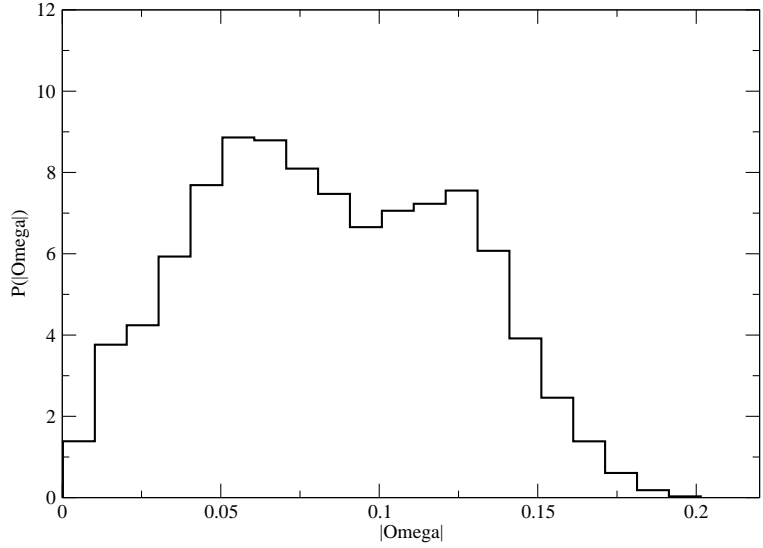}
\end{center}
\caption{Histogram for the modified probability distribution for a
 $6 \times 36^3$ lattice at $\beta_f=7.803$, $\beta_a=-2.0$.  
Details are given in the text.}\label{histo1}
\end{figure}

\begin{figure}
\begin{center}
\includegraphics[width=10cm]{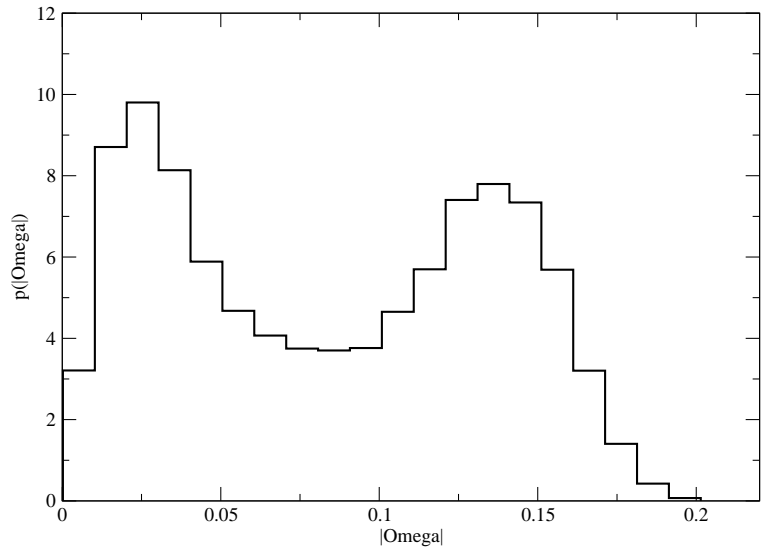}
\end{center}
\caption{Histogram for the Boltzmann distribution
 $6 \times 36^3$ lattice at $\beta_f=7.803$, $\beta_a=-2.0$. 
 Details are given in the text.}\label{histo2}
 \end{figure}

In fig. \ref{histo1} 
we give the histogram for $\exp(-S[U]) \; W(|\Omega|)$.
Indeed, with our choice of parameters for $W(|\Omega|)$, 
no deep minimum, separating the peaks, is visible. Fig. \ref{histo2} gives
the 
Boltzmann-distribution for $\beta_f=7.803$ and $\beta_a=-2.0$
obtained from the simulation. Now we see a clear minimum that separates
the peaks for the disordered phase and the three ordered phases. 
Still the weight of the two peaks is roughly the same. Therefore we 
performed a reweighting following eq.~(\ref{reweighting}), such that
eq.~(\ref{defbetac}) is satisfied. Under reweighting, the position of the 
minimum $O_{min}$ slightly shifts. To get a consistent result, we 
therefore repeat the analysis with the new value of $O_{min}$. We found 
that the procedure quickly converges, such that we get an estimate 
of $\beta_{f,c}$ with the proper $O_{min}$. The analysis of the statistical 
error of $\beta_{f,c}$ is done with the standard Jacknife procedure.


 \begin{figure}
 \begin{center}
 \includegraphics[width=10cm]{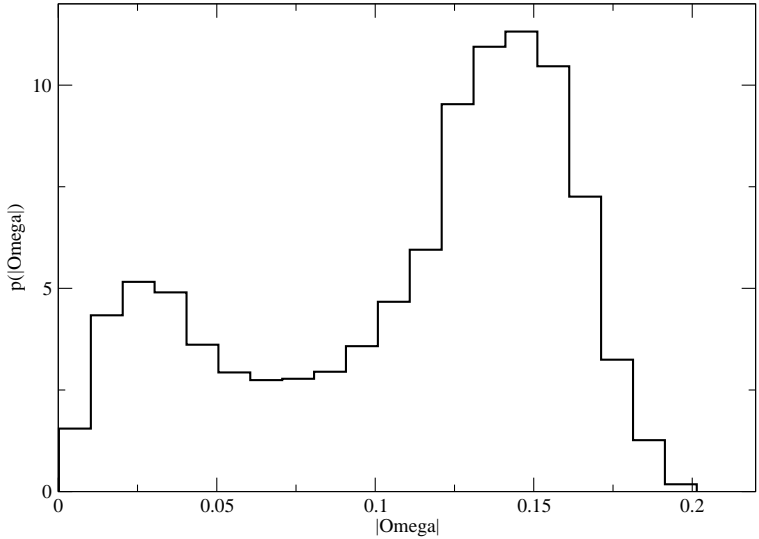}
 \end{center}
 \caption{Histogram for the proper probability distribution
 $6 \times 36^3$ lattice, simulated at $\beta_f=7.803$, $\beta_a=-2.0$,
 reweighted to $\beta_f=7.8056$.  Details are given in the text.}\label{histo3}
\end{figure}

\begin{table}
\caption{\sl \label{ourTc}
Summary of 
our results for the finite temperature phase transition obtained for
the fundamental-adjoint gauge action. We give the value of $\beta_{f,c}$ 
for fixed values of $N_t$ and $\beta_a$.  For a discussion of the numbers
see the text.
}
\begin{center}
\begin{tabular}{|l||l|l|l|}
\hline
 $N_t$;  $\beta_a$ &\phantom{00} 0.0&\phantom{0}--2.0 &\phantom{0}--4.0 \\
\hline
\hline
\phantom{0}  2   & 5.0948(6) &  6.4475(6)  &  7.8477(6) \\
\hline
\phantom{0}  3   & 5.5420(3) &  7.1603(3)  &  8.8357(4) \\
\hline
\phantom{0}  4   & 5.6926(2) &  7.4433(3)  &  9.2552(6) \\
\hline
\phantom{0}  6   & \phantom{000}-  & 7.8056(5) & 9.7748(11) \\
\hline
\end{tabular}
\end{center}
\end{table}

\begin{table}
\caption{\sl \label{otherTc}
Results for the finite temperature phase transition obtained by other
authors for the standard Wilson gauge action, i.e. $\beta_a=0$.
}
\begin{center}
\begin{tabular}{|r|l|l|}
\hline
 $N_t$&  $\beta_{f,c}$  & ref. \\
\hline
  2   &  5.0933(7)  & \cite{AlBeSa92} \\
\hline
  3   & -- & -- \\
\hline
  4   &   5.6927(4) &  \cite{AlBeSa92} \\
  4   &   5.69254(24)& \cite{QCDPAX} \\
  4   &   5.6925(2) & \cite{Bielefeld96} \\
\hline
  6   &   5.89405(51)& \cite{QCDPAX} \\
  6   &   5.8941(5) & \cite{Bielefeld96} \\
\hline
  8   &   6.0624(9)(3) & \cite{Bielefeld96},\cite{Beinlich:1997ia}\\
\hline  
 12   &   6.3380(13)(10) & \cite{Bielefeld96},\cite{Beinlich:1997ia}\\
\hline
\end{tabular}
\end{center}
\end{table}

Our results are summarized in Table \ref{ourTc}. For comparison, 
we have collected results for $\beta_a=0$ from the literature. In contrast
to our study, 
these results where obtained from the peak of the susceptibility.
In the two cases, where we have mapping parameters, our results 
are consistent with those of the literature. For 
$\beta_a=0$, $N_t =6$ we performed no own simulation, but used in the following 
the result $\beta_f = 5.89405(51)$ of ref. \cite{QCDPAX}.


\subsection{Lines of constant physics}
\label{linesof}
At first order in perturbation theory the tree-level relation 
(\ref{constant_tree_level}) is modified as
\cite{GoKo}:
\begin{equation}\label{constant_physics}
\beta_{W}=\beta_{f}+2\beta_{a}-5\frac{\beta_{a}}{\beta_{f}+2\beta_{a}} \;\;.
\end{equation}
Up
to higher order perturbative corrections and non-perturbative contributions,
lines of constant $\beta_{W}$ should represent 
\emph{lines of constant physics}.
\begin{figure}
\begin{center}
\includegraphics[width=10cm]{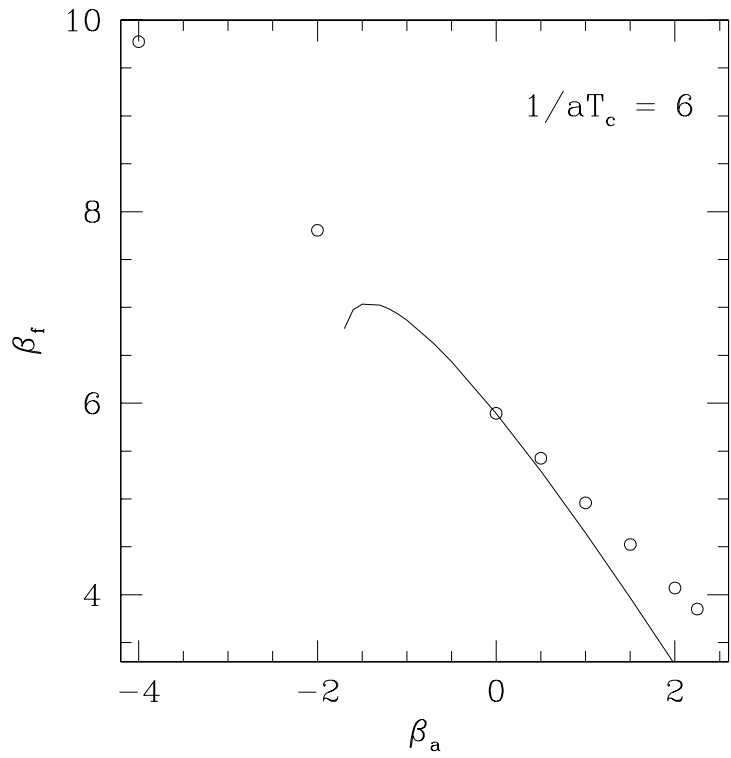}
\end{center}
\caption{Line of constant physics (solid line)
as predicted by eq.~(\ref{constant_physics})
and the deconfinement transition line for $a=0.11\,\rm{fm}$ (open circles).
}\label{coup}
\end{figure}
As a check of this relation, we compare in fig.  \ref{coup}
the values of the critical couplings for $N_t=6$, 
i.e. $a\simeq 0.11\,\rm{fm}$, 
\footnote{In order to express the lattice spacing in physical units, we
used the relation $T_c r_0=0.7498(50)$ \cite{SN03} and $r_0\simeq
0.5\,\rm{fm}$ \cite{sommer_r0}.} with
the one-loop prediction eq.~(\ref{constant_physics}), where we have set
$\beta_W=\beta_{f,c}|_{a=0.11\,\rm{fm},\; \beta_a=0}$.
For $\beta_a>0$ we adopt the results obtained in \cite{Blum:1995xb}.\\
We see that the perturbative prediction fails in describing the line
of constant physics. In ref. \cite{Blum:1995xb} 
it was observed that the prediction
for positive $\beta_{a}$ can be considerably improved by using a tadpole
improved perturbative formula. For negative $\beta_{a}$, however, the
perturbative prediction seems to be completely unreliable. Hence we made 
no attempt to study tadpole improvement here.

\begin{figure}
\begin{center}
\includegraphics[width=10cm]{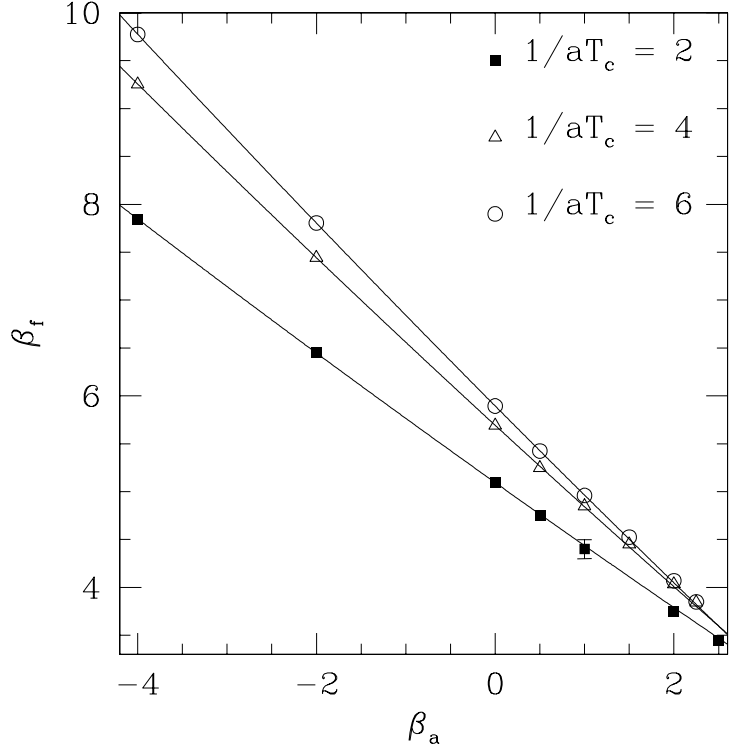}
\end{center}
\caption{Deconfinement transition lines for 
$a=0.11,0.17,0.33\,\rm{fm}$ and the interpolation
formulas eq.~(\ref{fit_constant}) (solid lines). 
For $\beta_{a}>0$ the data are taken 
from ref. \cite{Blum:1995xb}.}\label{coup_all}
\end{figure}

However, it turns out that the deconfinement transition lines can be 
parametrised quite well by a quadratic interpolation.
Fitting the full range of available $\beta_a$ (including the results
of ref. \cite{Blum:1995xb} for $\beta_a>0$) yields:
\begin{eqnarray}\label{fit_constant}
a=0.33\,\rm{fm}\,:\, \beta_f & = & 5.0948-0.6645\beta_a+ 0.0059\beta_a^2\\
a=0.17\,\rm{fm}\,:\, \beta_f & = & 5.6930-0.8586\beta_a+ 0.0080\beta_a^2\nonumber\\
a=0.11\,\rm{fm}\,:\, \beta_f & = & 5.8951-0.9391\beta_a +0.0078\beta_a^2\nonumber
\;\;.
\end{eqnarray}
Fig. \ref{coup_all} shows the deconfinement transition lines for 
$a=0.11,0.17,0.33\,\rm{fm}$ for a range of positive and negative $\beta_a$.
For $a=0.33\,\rm{fm}$, the interpolation formula reproduces the numerical
values for $\beta_a<0$ within the statistical errors.
For $a=0.17,0.11\,\rm{fm}$ and $\beta_a\leq 0$ the quadratic interpolation
reproduced the numerical data only within two standard deviations.


\section{The static potential}
We have extracted the static potential
from Polyakov loop correlation functions:
\begin{equation}
a V(r)=-\frac{1}{N_t} \left[ \ln\langle P(x)^* P(y)\rangle +\epsilon \right]
\;\;,
\end{equation}
where $y=x+r\hat{1}$.
$\epsilon$ is the correction due to excited states in the string spectrum.
It vanishes exponentially as $N_t \rightarrow \infty$.
In the free bosonic string approximation one gets 
\cite{minami,df83,flensburg,fsst}
\begin{equation}
\label{correction}
\epsilon=
2 \; \sum_{n=1}^\infty \ln (1-\exp[-\pi n a N_t/r]) \;\;.
\end{equation}
Note that here, in contrast to the previous section,
we are considering  systems  with a temporal extension $a N_t >> 1/T_c$ to
eliminate finite temperature corrections $\epsilon$
as much as possible. In the simulations reported below, we have chosen
$N_t = 6/(a T_c)$. It is straightforward to check that for this choice,
at our level of numerical precision, the corrections eq.~(\ref{correction})
can be safely ignored.

We have computed the Polyakov loop correlation function with a variant 
of the algorithm that was recently proposed by L\"uscher and Weisz \cite{lw}.
Details of the algorithm are given below.

In this study, we consider rather coarse lattice spacings. 
Therefore the Sommer scale  $r_0$  \cite{sommer_r0}  is
intrinsically affected by large systematic errors. On the other hand, we have
computed the static potential up to rather large physical distances $r$.
Therefore we decided to compute the string tension $a^2 \sigma$ rather than
$r_0/a$.
We evaluated the string tension using the  
ansatz 
\begin{equation}
\label{pot}
V(r)=\sigma r + \mu - \frac{\pi}{12 r}\left(1+\frac{b}{r}\right)
\end{equation}
for the static quark potential, 
with $b=0.04 {\rm fm}$ \cite{lw}, where contributions of the
order $O(r^{-3})$ are neglected. Details of our numerical analysis 
are given in the subsection \ref{stringsection}.

\subsection{Variant of the L\"uscher and Weisz method to compute the 
Polyakov loop correlation function}
In order to reach large values
of $N_t$ we have implemented a variant of the recent proposal 
of L\"uscher and Weisz
\cite{lw}.  In addition to the factorisation in temporal direction, 
we also use a factorisation in the spacial directions 
\footnote{In ref. \cite{Laine04} the Polyakov loop correlation function 
was measured with a spatial decomposition only. The model studied in this 
ref. contains scalar fields in addition to the gauge field}.

In temporal direction, we have used only one level of the factorisation.
To this end, the lattice is divided in temporal direction 
into $N_t/N_l$ layers of the thickness $N_l$. 
Note that L\"uscher and Weisz
\cite{lw} also have used only one level of the factorisation in most of their 
numerical studies.

\begin{figure}
\begin{center}
\includegraphics[width=10cm]{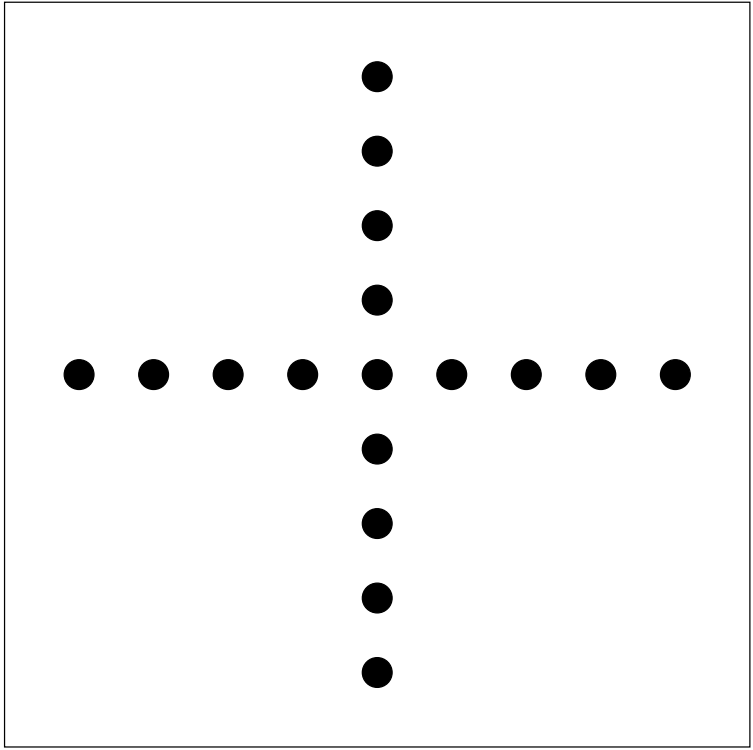}
\end{center}
\caption{{Two dimensional sketch of the block. Two of the 
spatial direction  are show. The remaining spatial direction and the 
the temporal direction are perpendicular to the paper. Only the Polyakov loops
that are indicated with black discs are computed.}}\label{block}
\end{figure}

In addition, we have used a factorisation in the spacial directions.
To this end, we have divided the lattice in blocks of the size 
$b^3 \times N_l$. Within these blocks we consider only a subset 
of all possible Polyakov loops. See the two-dimensional sketch fig. 
\ref{block} for an illustration. 
The idea of this choice is to use, for a given distance of the 
loops, only those loops that have
a maximal distance from the boundaries of the blocks.
In particular, 
for an even distance $r/a$ between the loops, we 
took the two loops with distance $(r/a)/2$ from the common boundary.
For an odd distance $r/a$ between the loops, we took the distances 
$(r/a+1)/2$ and $(r/a-1)/2$ from the boundary between the blocks.

We have now a two-fold hierarchy of the algorithm. At the lowest level, 
we update at fixed boundaries of the blocks and fixed boundaries between 
the temporal slices. This step provides us with variance 
reduced segments of the Polyakov loops
\begin{equation}
\label{segment}
\bar{P}(\vec{x},t_0) = \frac{1}{M_{block}}  \sum_{i=1}^{M_{block}}
                   \prod_{t=t_0}^{t_0+N_l-1} U_{\vec{x},t,0}^{(i)}
\;\;,
\end{equation}
where $M_{block}$ is the number of updates that have been performed 
with fixed boundaries of the block and $i$ labels the configurations
that have been generated this way.
These variance reduced segments of the Polyakov loop
could be viewed as a generalization
of the multi-hit method for the variance reduction of a single link 
variable that has been applied in ref. \cite{lw}.

Two of these variance reduced segments with the same $t_0$ from 
neighbouring spatial blocks are now used to construct the complex $9 \times 9$ 
matrices of eq.~(3.2) of ref. \cite{lw}: 
\begin{equation}
\label{Tmatrix}
 \mathbb{T}(\vec{x},\vec{y},t_0)_{\alpha,\beta,\gamma,\delta}
= \bar{P}(\vec{x},t_0)^*_{\alpha,\beta} 
  \bar{P}(\vec{y},t_0)_{\gamma,\delta} \;\;.
\end{equation}

For fixed boundaries between the temporal layers, we perform a certain 
number of update sweeps before we repeat again the calculation of the 
variance reduced segments eq.~(\ref{segment}). This procedure is performed
$M_{layer}$ times and the matrix $\mathbb{T}$ is averaged over these $M_{layer}$
instances:
\begin{equation}
\bar{\mathbb{T}}(\vec{x},\vec{y},t_0)_{\alpha,\beta,\gamma,\delta}
 = \frac{1}{M_{layer}}  \sum_{j=1}^{M_{layer}}
\mathbb{T}(\vec{x},\vec{y},t_0)_{\alpha,\beta,\gamma,\delta}^{(j)}
\;\;.
\end{equation}
As in ref. \cite{lw}, the Polyakov loop correlation function is now
computed as
\begin{equation}
\langle P(x)^* P(y)  \rangle \approx \frac{1}{M_{meas}} 
\sum_{k=1}^{M_{meas}}
\left\{\prod_{t=0}^{t=N_t/N_l-1}
\bar{\mathbb{T}}(\vec{x},\vec{y},t N_l)^{(k)}
\right\}_{\alpha,\alpha,\gamma,\gamma}
\;\;,
\end{equation}
where $M_{meas}$ is the number of complete measurement cycles.
It is understood that measurements start after equilibration of the system.
 
In order to further improve the measurement of the segments of the Polyakov
loop
eq.~(\ref{segment}) we have performed a multi-hit update of the single 
link variables. In addition, we have updated the links close
to the centre of the block more frequently than those close to the 
boundary.  To this end, we have set up a sequence of blocks inside
the block having the sizes $(b-2)^3$, $(b-4)^3$, ... .
For each cycle (that consists of 2 OV-Metropolis and 2 CM-Metropolis sweeps
over a block of the spatial size $(b-2m)^3$) we have performed 
$n$ such cycles 
for the next smaller block $(b-2(m-1))^3$. I.e. for
each cycle of the full block ($b^3$),
 additional $n$, $n^2$, $n^3$... cycles are 
performed for the blocks of the spatial size 
$(b-2)^3$, $(b-4)^3$, $(b-6)^3$, ...
inside the block of the spatial size $b^3$. 
In most of our simulations we have chosen $n=3$.
This way, of course, we can not gain 
exponentially, but we can fight, to some extend, the factor that we loose 
by the reduced number of Polyakov loops that we consider.


Compared with the original approach of L\"uscher and Weisz, we gain the 
factorisation in the spacial directions. However, we lose a lot of copies 
of the correlator. Instead of $3 \times N_s^3$ only $3 \times (N_s/b)^3$ remain.
To compensate partially for this fact, we update more frequently the links 
in the centre of the block than those at the boundary.
A positive side-effect of the reduced number of copies of the Polyakov loop
is that the memory requirements are
drastically reduced compared with the original proposal.

Most of the CPU-time is spent with the block-boundaries fixed. This 
would allow for a rather trivial parallelisation on  $(N_t/N_l) (N_s/b)^3$
nodes. This makes the algorithm ideally suited for a PC-cluster 
equipped with a moderately fast interconnect like Gigabit-ethernet.

The obvious disadvantage of our variant of the algorithm is that even 
more parameters have to be tuned as in the original one of L\"uscher
and Weisz. In our case, it 
is almost unavoidable to make just ad hoc choices for 
some of the parameters.   Finally, for the distances $r/a$ that we 
have reached, our method still seems to be slightly less efficient than 
the one of ref. \cite{lw}. Here is difficult to give a definite comparison, 
since we did not simulate at exactly the same  parameters as ref. \cite{lw}.

\begin{table}
\caption{\sl \label{sim_det}
Parameters of our simulations to compute the static quark potential. 
In the first two columns we give the coupling constants. The third
column contains the lattice sizes. In the fourth column the range 
of the distance $r$ is shown where the static potential has been
computed. Finally we give the number of measurements $M_{meas}$,
as defined in the previous section. 
}
\begin{center}
\begin{tabular}{|l|l|l|l|c|}
\hline
\multicolumn{1}{|c}{$\beta_a$} & 
\multicolumn{1}{|c}{$\beta_f$}  & 
\multicolumn{1}{|c}{$N_s$, $N_t$}  &  
\multicolumn{1}{|c}{$r/a$}   & 
\multicolumn{1}{|c|}{$M_{meas}$} \\
\hline
\phantom{--}0 &  5.0848          &  16, 12  &  2-6        &  30  \\ 
        --2   &  6.4475          &  16, 12  &  2-6       &  26  \\
        --4   &  7.8477          &  16, 12  &  2-6      &  24  \\
\hline   
\phantom{--}0 &  5.5420          & 20, 18  &  2-8         &  17  \\
        --2    &  7.1603          & 20, 18  &  2-8         &  17  \\
        --4    &  8.8357          & 20, 18  &  2-8         &  21  \\
\hline   
 \phantom{--}0&  5.6926           & 24, 24  &  2-10        &  56  \\
        --2   &  7.4433           & 18, 24  &  2-4         &  76  \\
        --4   &  9.2564           & 18, 24  &  2-4         &  101\phantom{0} \\
        --4   &  9.254            & 18, 24  &  2-4         &  76 \\
\hline
\end{tabular}
\end{center}
\end{table}
\subsection{Numerical results for the string tension}
\label{stringsection}
To eliminate the constant $\mu$ in eq.~(\ref{pot})
we consider the so called force 
\begin{equation}
\label{force}
F(r) = \frac{\mbox{d} V}{\mbox{d} r} = \sigma 
+ \frac{\pi}{12 r^2}\left(1+\frac{2 b}{r}\right) \;\;.
\end{equation}
On the lattice, we compute the force 
from the potential either as
\begin{equation}
a F(r-a/2) = V(r)-V(r-a)
\label{naive}
\end{equation}
or
\begin{equation}
a F(r_I) = V(r)-V(r-a) \;\;,
\label{improved}
\end{equation}
where $r_I$ is the tree-level improved distance defined in ref. 
\cite{sommer_r0}.

In our numerical analysis, we made no attempt to compute $b$, but instead
have used the result $b=0.04 {\rm fm}$ of ref. \cite{lw}. In order
to obtain the dimensionless quantity $a^{-1} b$ we have used
$T_c r_0 =0.7498(50)$ obtained in ref. \cite{SN03} with $r_0 =0.5 {\rm fm}$.
It follows
\begin{equation}
\label{am1b}
a(\{\beta_f,\beta_a\}_c )^{-1} b \approx 0.06 N_t \;\;,
\end{equation}
where here $a(\{\beta_f,\beta_a\}_c ) N_t =1/T_c$.
As a result, eq.~(\ref{force}) has no free parameters in addition to $\sigma$.
Hence, for each value of $r/a$ we obtain an estimate for $a^2 \sigma$.

We computed the static quark potential at the critical couplings that we have
evaluated in the previous section for
$\frac{1}{T_c a}=2,3,4$. 
The parameters of these simulations are given in Table \ref{sim_det}.
As thickness of the temporal layers we have used $N_l=1/(a T_c)$ throughout.
For the other parameters, let us just detail a typical example:
For $\beta_a=0$, $\beta_f=5.6926$ we have used $M_{layer}=30$ and 
$M_{block}=40$.

In Table \ref{sigmaNt4} we give details of our analysis of the force
for $\beta_a=0$, $\beta_f=5.6926$, where we have collected our 
most accurate data. First we have estimated the string tension as
$a (V(r)-V(r-a))$, ignoring the L\"uscher term. We see that even for 
our largest values of $r/a$ the estimate does not stabilize. This behaviour
is clearly improved, when the L\"uscher term is used in the ansatz.
Fortunately, the difference between the two definitions of the argument 
of the force eq.s~(\ref{naive},\ref{improved}) is only minor. Finally, 
we included $b=0.04 {\rm fm}$ in the ansatz. As a result, the estimate 
of the string tension further stabilizes as the distance $r/a$ is varied.
In fact, starting from $r/a=4$, all results are consistent within the 
statistical error.

As our final result we quote $a^2 \sigma=0.162(1)$, which is obtained from 
the ansatz with $r_I$ and $b=0.04 {\rm fm}$ at the distance $r/a=7$.
Since the difference of our results with $b=0$ and $b=0.04 {\rm fm}$
at $r/a=7$ is smaller than the statistical error, 
we are confident that the quoted error also covers possible systematic errors.

\begin{table}
\caption{\sl \label{sigmaNt4}
Analysis of the force at $\beta_a=0$, $\beta_f=5.6926$.
In this Table we give results obtained with various ansatze. In the second
column we give the most naive estimate for the string tension $a^2 \sigma$:
$a (V(r)-V(r-a))$. In the third, we use the ansatz~(\ref{force}) in connection
with eq.~(\ref{naive}) and $b=0$. In the fourth column, the naive definition
of the distance is replaced by the improved one eq.~(\ref{improved}), 
still $b=0$. Finally, in the
column 5 we use $b=0.04 {\rm fm}$ as discussed in the text. 
In the last column we
report the statistical error, which is the same in all cases.
}
\begin{center}
\begin{tabular}{|c|c|c|c|c|c|}
\hline
 $r/a$ & $a^2 \sigma_{naive}$  & $a^2 \sigma_{b=0}$
     &  $a^2 \sigma_{b=0}$,$r_I$  &   $a^2 
   \sigma_{b=0.04 {\rm fm}}$,$r_I$ &  stat. error \\
\hline
 3 & 0.2211 & 0.1793 & 0.1707 & 0.1600 & 0.0003 \\
 4 & 0.1902 & 0.1688 & 0.1664 & 0.1629 & 0.0004 \\
 5 & 0.1783 & 0.1654 & 0.1645 & 0.1630 & 0.0004 \\
 6 & 0.1727 & 0.1640 & 0.1637 & 0.1629 & 0.0006 \\
 7 & 0.1692 & 0.1630 & 0.1628 & 0.1623 & 0.0006 \\
 8 & 0.1667 & 0.1621 & 0.1620 & 0.1617 & 0.0011 \\
 9 & 0.1655 & 0.1618 & 0.1618 & 0.1616 & 0.0025 \\
\hline
\end{tabular}
\end{center}
\end{table}

We have extracted the final result for $a^2 \sigma$ for the other values
of the couplings in a similar fashion. In particular, we have always taken 
the result obtained with the 
tree-level improved distance $r_I$  and with $b=0.04{\rm fm}$. 
For $1/(a T_c)=2$ and $3$, the final results for $a^2 \sigma$ are taken
from $r/a=4$ and $r/a=6$, respectively.
In the case of $1/(a T_c)=4$ we give
the final result for $\beta_a=0$ and $\beta_f=5.6926$ that we obtained
above. For $\beta_a=-2$ and $\beta_a=-4$ we do not have data for such
large distances. Therefore we took $r/a = 4$. The error that is quoted
here includes a systematic error that we estimate based on our data
for $\beta_a=0$ and $\beta_f=5.6926$.  These results are  summarized
in Table \ref{sigma_table}.

M. L\"uscher \cite{luescher_private} has provided us with numerical
data for the force 
at $\beta_f=5.7$, $\beta_a=0.0$ obtained on a $24 \times 18^3$ lattice 
\cite{lw}.
Using the procedure discussed above, we extract $a^2 \sigma = 0.156(1)$
from these data, where 
the error mainly covers possible systematic errors. (The statistical error 
of the force at $r/a=7$ is only $1.5 \times 10^{-4} $). Extrapolating our
data for $\beta_f=5.5420$, $\beta_a=0$ and $\beta_f=5.6926$, $\beta_a=0$,
using the ansatz $\ln(a^2 \sigma) = c + d \beta$, we get  
$a^2 \sigma = 0.1567(10)$ for $\beta_f=5.7$, $\beta_a=0$, which is consistent
with the result that we have extracted from the numerical data of 
ref. \cite{lw,luescher_private}. 

In Table \ref{sigma_table} we also give the results for $T_{c}/\sqrt{\sigma}$, 
which are plotted in
fig. \ref{sigma_table} together with other values obtained for the Wilson 
action in \cite{Beinlich:1997ia}. 
The error of $T_{c}/\sqrt{\sigma}$ is dominated by
the error of $a^2 \sigma$.

\begin{table}
\caption{\sl \label{sigma_table}
Summary of our final results for the string tension 
$a^2 \sigma$ and the dimensionless
ratio $T_c/\sqrt{\sigma}$. For a detailed discussion see the text.
}
\begin{center}
\begin{tabular}{|l|l|c|c|c|}
\hline
$1/(a T_c)$ & $\beta_a$ & $\beta_f$  &  $a^{2} \sigma$ & $T_c/\sqrt{\sigma}$ \\
\hline
 2 & \phantom{--}0  &  5.0948 &  0.759(2)  &  0.574(1)\\
   &        --2  &  6.4475 &  0.742(2)  &  0.580(1)\\
   &        --4  &  7.8477 &  0.736(2)  &  0.583(1)\\
\hline
 3 & \phantom{--}0  & 5.5420   &  0.319(2) & 0.590(2) \\
   &       --2  & 7.1603   &  0.308(2) & 0.601(2) \\
   &       --4  & 8.8357   &  0.306(2) & 0.603(2) \\
\hline
4  & \phantom{--}0  &  5.6926   &   0.162(1) &  0.621(2) \\
   &      --2  &  7.4433   &   0.160(1) &  0.625(2) \\
   &      --4  &  9.2540   &   0.160(1) &  \\
   &      --4  &  9.2564   &   0.159(1) &  \\
   &      --4  &  9.2552    &            & 0.626(2) \\
\hline
\end{tabular}
\end{center}
\end{table}

We see that for $\beta_a=-2$ and $-4$ the estimate for $T_c/\sqrt{\sigma}$
is closer to the continuum limit than for $\beta_a=0$.  However,
the difference between $1/(a T_c)=3$ and $1/(a T_c)=4$ is larger than
that for the different values of $\beta_a$ at fixed $1/(a T_c)$.
Since already for $1/(a T_c)=4$ there is only a minor difference 
in the estimates for $T_c/\sqrt{\sigma}$ from $\beta_a=0$ 
and $\beta_a=-2$,$-4$, we decided
to skip the determination of $a^2 \sigma$ for $1/(a T_c)=6$.

\begin{figure}
\begin{center}
\includegraphics[width=10cm]{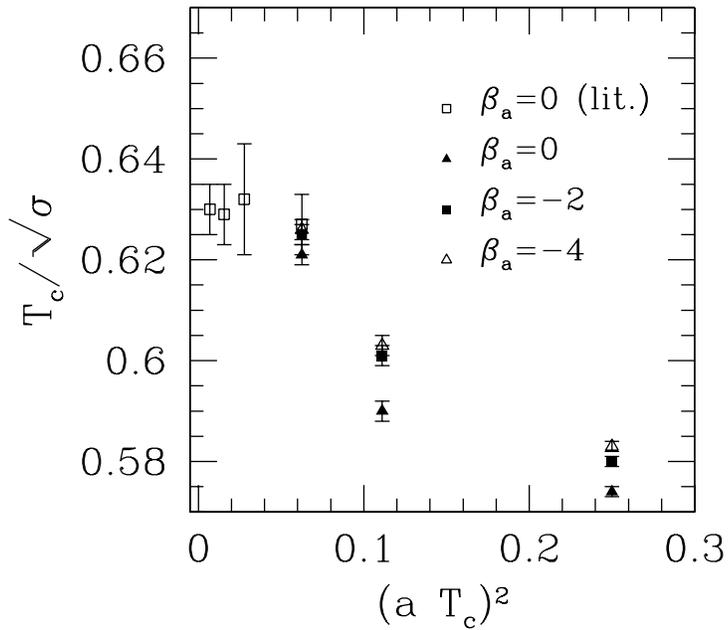}
\end{center}
\caption{We have plotted our results for $T_c/\sqrt{\sigma}$ as a function 
of $1/N_t^2$, where $N_t=1/(a T_c)$. In addition we give the results obtained 
in ref. \cite{Beinlich:1997ia} 
for $\beta_a=0$ at smaller lattice spacings.}\label{fig_string}
\end{figure}

For comparison, we have summarised in Table \ref{tab_string_cont}
continuum limit results for 
$T_{c}/\sqrt{\sigma}$ given in the literature that are 
obtained with various actions.  
In particular, the result of ref.s \cite{Beinlich:1997ia} and \cite{Bliss}
are not compatible within the given error bars. Given that our smallest
lattice spacing is $a \approx 0.17$ fm, we did not extrapolate our data to 
the continuum limit. However, 
looking at fig. \ref{fig_string}, our data seem to be more compatible with
a continuum result $T_{c}/\sqrt{\sigma} \approx 0.66$ 
than $0.63$. 

\begin{table}
\caption{\sl {Results for the continuum limit of $T_{c}/\sqrt{\sigma}$
 obtained with various lattice actions.}}\label{tab_string_cont}
\begin{center}
\begin{tabular}{|l|l|}
\hline
action & $T_{c}/\sqrt{\sigma}$\\
\hline
Wilson \cite{Beinlich:1997ia}        & 0.630(5)\\
Symanzik imp. \cite{Beinlich:1997ia} & 0.634(8)\\
DBW2  \cite{deForcrand:1999bi}       & 0.627(12)\\
Iwasaki \cite{Okamoto:1999hi}         & 0.651(12)\\
1-loop tadpole impr. \cite{Bliss}     & 0.659(8) \\
\hline
\end{tabular}       
\end{center}
\end{table}

\newpage

\section{Glueball masses}
\begin{table}
\caption{\sl Parameters of the glueball computation. For definitions 
see the text.}\label{sim_par}
\begin{center}
\begin{tabular}{|c c c r r c c r c r r|}
\hline
$\beta_{f}$ &  $\beta'_{f}$ & $\beta_{a}$ &  $N_s$ & $N_t$ & $n_{l}$ & $N_{or}$ &  $N_{sub}$  & $I_{sub}$ & $I_{gl}$ & $N_{meas}$ \\
\hline
5.0948 &   - &  \phantom{--}0\phantom{.0} &  4 & 6 & 1,2,3,4 & 2 &     320 &          4 &                6 & 9931\\
6.4475 &  4.5 & --2.0 &  4 & 6 & 1,2,3,4  &   2 &     160 &         2  &               6 & 19800 \\
7.8477 &  4.5 & --4.0 &  4 & 6 & 1,2,3,4 &   2 &      160 &         2  &               6 & 17800\\
\hline
5.5420 &   - &     \phantom{--}0\phantom{.0}  &   6 &  8 &  2,4,6,8 & 4 &     160 &          4 &                8 & 4894  \\
7.1603 &  5.0 &  --2.0 &  6 & 8 & 2,4,6,8  &   4 &      80 &         2  &               8 & 8500\\
8.8357 &  5.0 &  --4.0 &  6 & 8 & 2,4,6,8  &   4 &      80 &         2  &               8 & 7140\\
\hline
5.6926 &   - & \phantom{--}0\phantom{.0}  &   8 & 12 & 2,4,6,8 &    5 &     160 &          4  &               8 & 4520 \\
7.4433 &  5.0 &   --2.0 &  8 & 12 & 2,4,6,8  &   5  &     80  &        2    &             8 & 2925\\
9.2564 &  5.0 &   --4.0 &  8 & 12 & 2,4,6,8  &   5  &     80  &        2    &             8 &  3852  \\
\hline
\phantom{0}5.89405 
&  - &       0  &  12 & 18 & 3,6,9,12 &   7 &     300  &  6  &  10 & 584\\
7.8056  &  5.0 &    -2.0 &  12 & 18 & 3,6,9,12 &   7 &      100 &         2 &               10 & 672  \\
9.7748  &  5.0 &    -4.0 &  12 & 18 & 3,6,9,12  &  7  &    100  &        2  &              10 & 833\\
\hline
\end{tabular}
\end{center}
\end{table}
We continued our investigation of the lattice artefacts of the 
mixed action by  
measuring the mass (at zero temperature) 
of the lightest glueball $0^{++}$ at the
critical values of $\beta_{f}$ for $\beta_{a}=0,-2,-4$ and
$1/(a T_c)=2,3,4,6$ that we have determined in section \ref{finiteTsection}.
This allows us to
study the scaling of the dimensionless quantity $m_{0^{++}}/T_c$. 
The $0^{++}$ glueball mass is particularly interesting since it 
shows large lattice
artefacts in the case of the Wilson action ($\beta_a=0$).
The fact that the $0^{++}$ mass becomes very small at
certain lattice spacings can be interpreted as the influence of the endpoint
of the first order phase transition (see eq.~(\ref{endpoint})), where 
$m_{0^{++}}$ vanishes \cite{Heller}.
If this picture is correct, we expect that by choosing a negative $\beta_a$,
we move away  from the endpoint and hence the lattice artefacts 
on $m_{0^{++}}$ should be reduced.

In ref. \cite{Gupta91} an  action with 4 terms had been used: In addition 
to the plaquette in the fundamental representation, the plaquette in the 
adjoint representation, the representantion of dimension 6 and a $1 times 2$
loop in the fundamental representation were considered.  All couplings except 
that of the fundamental plaquette were negative. The authors argue
that this way the end-point of line of first order phase transtions 
in the fundamental-adjoint can be avoided.
Their final estimate for the mass of the lightest glueball is
$m_{0^{++}}/sqrt(\sigma)=3.5(3)$, which corresponds to 
$m_{0^{++}} r_0=4.07(35)$. 

On anisotropic lattices ($a_t << a_s$) the authors of ref.
\cite{MorningstarPeardon1} found that by introducing in the action a term 
$\mbox{Re} \mbox{Tr} U_P(\vec{x},t,\mu,\nu) \; 
\mbox{Re} \mbox{Tr} U_P(vec{x},t+a_t,\mu,\nu)$ for space-like plaquettes
with a  negative coupling constant, the scaling behaviour 
of the $0^{++}$ glueball mass can be improved. On top of this modification
they \cite{MorningstarPeardon2} employed Symanzik-improvement \cite{SYM,LuWe85} 
with a $2 \times 1$ Wilson loop term in the action.

\begin{figure}
\begin{center}
\includegraphics[width=8cm]{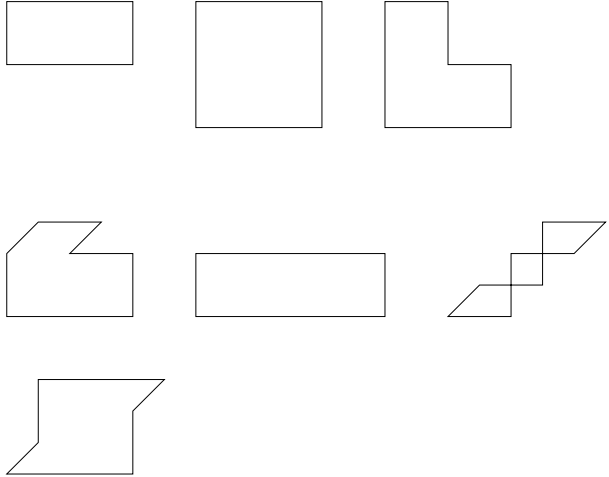}
\end{center}
\caption{Wilson loops used for the evaluation of $m_{0^{++}}$.}\label{glue_op}
\end{figure}

In order to extract the mass $m_{0^{++}}$ of the $0^{++}$ glueball,
we have computed the connected correlation function between spatial
Wilson loops. To this end,
we have chosen a basis of $N=7$ operators in the $A_{1}^{++}$ 
representation of the cubic group
(among the 22 spatial Wilson loops up to lenght 8) 
plotted in fig. \ref{glue_op}, 
which had shown the best signal-to-noise ratio 
in a previous study \cite{SN03}.

In order to
get a better overlap with light states, 
we applied the APE smearing procedure to the spatial links
\cite{smear:ape}. For each operator, 
we used $M=4$ different smearing levels.  
For each shape with $d$ different orientations, we measured the observable
\begin{equation}
O_{l}(t)=\frac{N_s^{-3/2}}{\sqrt{d}}\sum_{\vec{x}}\sum_{n=1}^{d}{\rm Tr} W_{l}^{n}(\vec{x},t) \;\;,
\end{equation}
where $W_{l}^{n}$ is the spatial Wilson loop with smearing level $l$, 
in the orientation $n$.
We then constructed the correlation matrix
\begin{equation}\label{glue_corrfu}
C_{ij}(t)=\langle O_{i}(t)O_{j}(0)\rangle-\langle O_{i}(t)\rangle\langle O_{j}(0)\rangle \;\;,
\end{equation}
where now the indices $i,j$ run from 1 to $N\times M=28$.  

\subsection{Factorization and error reduction}
The exponential error reduction  proposed by L\"uscher and Weisz \cite{lw}
and adopted in our work to compute Polyakov loop
correlation functions, can be applied to a wider class of $n$-point
functions. 
\footnote{A similar idea had been
exploited in ref. \cite{michael} to compute fermion propagators.}
In particular, this idea has been tested on the 
Wilson loop correlation functions to compute the scalar and tensor glueball 
masses \cite{meyer}.

Here we have implemented only one level of factorisation. 
To this end, 
we divided the lattice in temporal direction into two sub-lattices  with
an extension $\Delta_t= N_t/2$ each. 
While keeping
the spatial links at the time slices $t_{0}=1$ and
$t_{1}=N_t/2+1$ fixed \footnote{In this section, we have set $a=1$ to 
simplify the notation},
we 
performed $\mathcal{N}_{sub}$ ``sub-measurements''
\begin{equation}
\label{submeas}
 \bar {O}_i(t)   = \frac{1}{\mathcal{N}_{sub}}
                   \sum_{n=1}^{\mathcal{N}_{sub}} {O}_i(t)^{(n)} \;\;.
\end{equation}
For each of these measurements, we perform $I_{sub}$ sub-updating cycles.
Every updating consists of one Cabibbo-Marinari heatbath sweep and $N_{or}$
overrelaxation sweeps.
``Sub-updating'' means that all link-variables except those at $t_{0}$ and
$t_{1}$ are updated. The total number 
of sub-update cycles with one set of fixed link variables at 
$t_{0}$ and
$t_{1}$ is then $N_{sub} = I_{sub} \mathcal{N}_{sub}$.

After these $N_{sub}$ sub-update cycles with fixed links at 
$t_{0}$ and $t_{1}$,
we performed $I_{gl}$  global sweeps, with the same ratio $N_{or}$ between
overrelaxation and heatbath sweeps.
For $t$
even, we evaluated the 2-point ``unsubtracted'' correlation function as
\begin{equation}
C_{ij}^{u}(t) \approx \frac{1}{N_{meas}} \sum_{j=1}^{N_{meas}}
\frac{1}{2}\sum_{t'=1,N_t/2+1} \bar{O}_{i}(t'+t/2) \; \bar{O}_{j}(t'-t/2)
\end{equation}
while
for $t\geq 2$, $t$ odd, we used
\begin{eqnarray}
C_{ij}^{u}(t)
\approx \frac{1}{N_{meas}} \sum_{j=1}^{N_{meas}}
\frac{1}{4}\sum_{t'=1,N_t/2+1} 
 \bar{O}_{i}(t'+(t+1)/2) \; \bar{O}_{j}(t'-(t-1)/2) \;\;\; \nonumber \\
\;\;\;\;\;\;\;\;\;\;\;\;\;\;\;\;\;\;\;
+\; \bar{O}_{i}(t'+(t-1)/2) \; \bar{O}_{j}(t'-(t+1)/2)  \;\;,
\end{eqnarray}
where $N_{meas}$ is the number of times eq.~(\ref{submeas}) is evaluated.
An important point on the subtraction of the vacuum expectation value is that
only the measurements included in $C_{ij}^{u}(t)$ are taken 
into account. E.g. for $t$ even, we subtract
\begin{equation}
\frac{1}{N_{meas}^2} \left(\sum_{j=1}^{N_{meas}} 
 \bar{O}_{i}(t'+t/2)  \right)
\left( \sum_{j=1}^{N_{meas}} \bar{O}_{j}(t'-t/2) \right) \;\;.
\end{equation}
In this way, the two terms are highly statistically correlated, and hence
statistical fluctuations cancel, when the difference is taken. For a more
detailed discussion of this point see ref. \cite{meyer}.
Alternatively one could consider a temporal derivative of the
correlation function, as proposed in ref. \cite{Majumdar:2003xm}.

The free parameter of this 2-level scheme is the number of
sub-measurements $\mathcal{N}_{sub}$.  Here, we did not try to tune 
$\mathcal{N}_{sub}$ but rather followed the rule of ref.
\cite{meyer}:
\begin{equation}
\mathcal{N}_{sub}\simeq e^{m\overline{t}} \;\;,
\end{equation}
where $m$ is the mass of the state that should be measured and $\overline{t}$
is the time, 
where we intend to extract the mass from the
exponential decay of the 2-point function.

\begin{figure}
\begin{center}
\includegraphics[width=10cm]{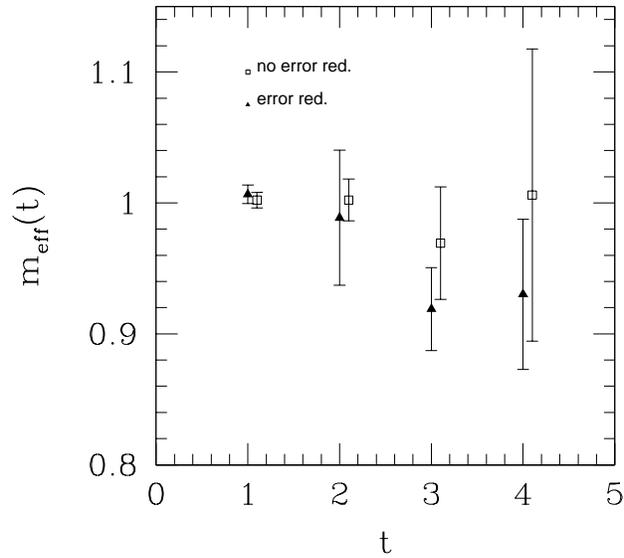}
\end{center}
\caption{The plot shows the effective mass of the  $0^{++}$ glueball 
at $\beta_{f}=5.6926$, $\beta_{a}=0$. In one case (squares) the correlation 
matrix has been calculated in the standard way, while in the other
case (triangles)   the variance reduction method that is discussed in the 
text has been applied. Note that for $t=1$ also the triangle result is 
computed in the standard way.
}\label{0_5.6926_red}
\end{figure}

In fig. \ref{0_5.6926_red} we analyse the effective mass evaluated with the
usual algorithm and with the factorisation formula for $\beta_{f}=5.6926$,
$\beta_{a}=0$, up to $t=4$. For this purpose we collected 16000 measurements
performed with the usual method and 400 measurements obtained with one level of
factorisation, where each measurement is obtained with 40
sub-measurements. The computational effort for these two simulations is then
roughly the same. The masses were extracted with the variational method
discussed below.
One can notice that for our choice of the algorithm, for
$t=2$ the standard method is still more efficient that the 
variance reducing one.
In our final analysis, 
we did not make use of the factorisation method
for time separations $t=0,1$
and we computed the correlation function in the usual way.
For $t\geq 3$ we observe however a substantial error reduction, and actually
this is the region where we extract the glueball mass. 

\subsection{Analysis and results}
\begin{figure}
\begin{center}
\includegraphics[width=10cm]{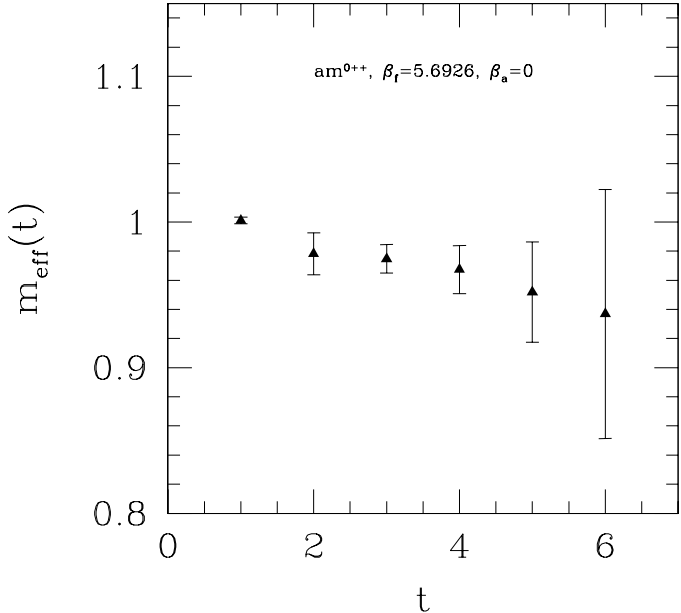}
\end{center}
\caption{Effective mass for the $0^{++}$ state, at $\beta_{f}=5.6926$, 
$\beta_{a}=0$ as a function of the distance $t$.}\label{0_5.6926}
\end{figure}

As a first step, we symmetrised the correlation matrix
by replacing $C_{ij}(t)$ by $1/2 \left[C_{ij}(t)+C_{ji}(t)\right]$.
The masses were extracted from the correlation functions by applying the
usual variational method \cite{pot:michael_SU3a,phaseshifts:LW}.
We solved the generalised eigenvalue problem
\begin{equation}
C(t) \; v_{\alpha}(t,t_{0})=\lambda_{\alpha}(t,t_{0}) \; C(t_{0}) \; 
v_{\alpha}(t,t_{0}) \;\;,
\end{equation}
with $\alpha=0,...,(N\times M)-1$ and
$\lambda_0>\lambda_1>...>\lambda_{(N\times M)-1}$. The indices $i,j$ of the
correlation matrix $C(t)$ are now omitted. 
In our analysis, we have chosen $t_{0}=0$ throughout.
We projected the correlation matrix
on the ground state eigenvector computed for $t=t_{0}+1$
\begin{equation}
W(t)=v_{0}^{T}(t_0+1,t_0) \; C(t) \; v_{0}(t_0+1,t_0) \;,
\end{equation}
and evaluated the effective mass using
\begin{equation}
m_{eff}(t)= \log\left(W(t-1)+\sqrt{W(t-1)^{2}-W(N_t/2)^{2}} \right) \;-
\end{equation}
$$
\log\left(W(t)+\sqrt{W(t)^{2}-W(N_t/2)^{2}}\right) \;,
$$
which takes into account the periodic boundary conditions in temporal 
direction.
In fig. \ref{0_5.6926} we show as an example the effective mass for
$\beta_f=5.6926$, $\beta_a=0$ as a function of the distance $t$. 
We decided to extract the mass at
$\overline{t}=4$, where the contributions of excited states should be
smaller than the statistical errors. (Note that the mass of the first 
exited state in the $A_1^{++}$ channel is almost twice as heavy as the $0^{++}$
state.)
Our final results are summarised in Table \ref{results_tab}, 
together with the value of $\overline{t}$, where we extracted the masses.

\begin{figure}
\begin{center}
\includegraphics[width=12cm]{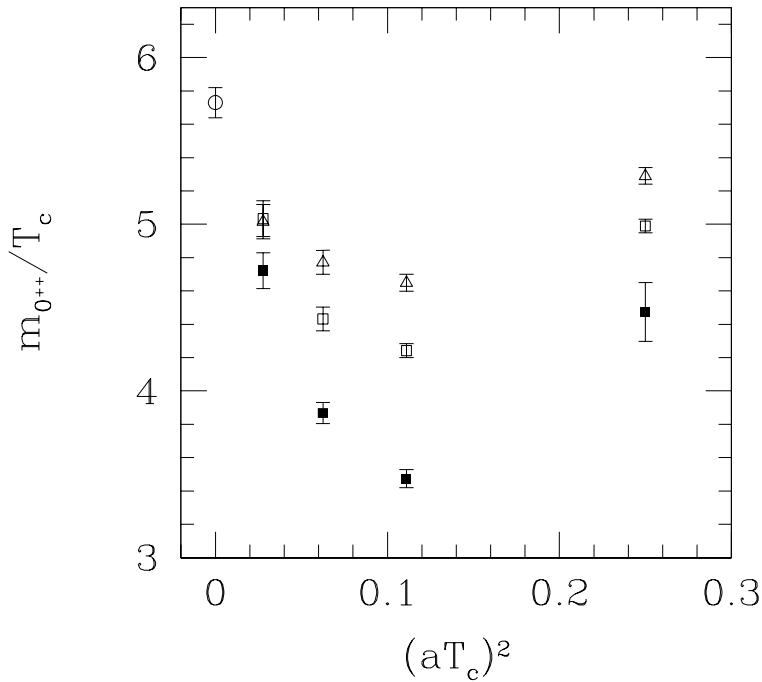}
\caption{$m_{0^{++}}/T_c$ for $\beta_a=0$ (filled squares) 
$\beta_a=-2$ (open squares) and $\beta_a=-4$ (triangles)
as function of $(aT_c)^2$. The circle 
gives the continuum results extracted from the literature.}\label{mass_scaling}
\end{center}
\end{figure}

\begin{table}
\caption{\sl Results for the $0^{++}$ glueball mass in lattice units. 
As our final result, we have taken the effective mass evaluated 
at the distance $\overline{t}$.
}\label{results_tab}
\begin{center}
\begin{tabular}{|c|c|c|c|l|}
\hline
$\beta_{f}$ & $\beta_{a}$ & $\overline{t}$ & $a m_{0^{++}}$& $m_{0^{++}}/T_c$\\
\hline
5.0948 & \phantom{--}0\phantom{.0} & 3   & 2.237(88) &  4.47(18)   \\
6.4475 & --2.0 & 2 & 2.495(20) &  4.990(40)  \\
7.8477 & --4.0 & 2 & 2.645(25) &  5.290(50)  \\
\hline
5.5420 & \phantom{--}0\phantom{.0} & 3   & 1.158(18) &  3.474(54)   \\
7.1603 & --2.0 & 2 & 1.414(14) &  4.242(42)   \\
8.8357 & --4.0 & 2 & 1.550(17) &  4.650(51)  \\
\hline
5.6926 &   \phantom{--}0\phantom{.0} & 4 & 0.967(16) &  3.868(64)  \\ 
7.4433 & --2.0 & 3 & 1.108(18) &  4.432(72)  \\
9.2564 & --4.0 & 3 & 1.193(18) &  4.772(72)  \\
\hline
 \phantom{0}5.89405 & 4 &  \phantom{--}0\phantom{.0}  & 0.787(18) &  4.72(11) \\
7.8056 & --2.0 & 3 & 0.839(18) &  5.03(11)   \\
9.7748 & --4.0 & 3 & 0.836(17) &  5.02(10)  \\
\hline
\end{tabular}
\end{center}
\end{table}
Our final results for $m_{0^{++}}/T_c$ are also reported in Table
\ref{results_tab}. Here the dominant error is the uncertainty on $m_{0^{++}}$.
Note that we computed the glueball mass for $\beta_f=9.2564,\beta_a=-4.0$,
which  was a preliminary estimate for the critical coupling
$\beta_{f,c}$
and not the final result
reported in Table \ref{ourTc}. However, we estimate that the
shift in the $0^{++}$ glueball mass corresponding to the shift in
$\beta_f$ is negligible with respect to the statistical error.
Fig. \ref{mass_scaling} shows $m_{0^{++}}/T_c$ 
as function of $(aT_c)^2$. 

For comparison, we 
give an estimate of the continuum limit based on
the average of the following results
given in the literature:
\begin{center}
\begin{tabular}{|l c|}
\hline
$m_{0^{++}}r_0$  & ref.\\
\hline
4.35(11)    &  \cite{glueb:teper98}\\
4.33(10)    &  \cite{Vaccarino:1999ku}\\
4.21(11)(4) &  \cite{Morningstar:1999rf}\\
4.23(22)    &  \cite{Liu:2001gx}\\
\hline
4.30(6)    &  average\\
\hline
\end{tabular}
\end{center}
The computation presented in \cite{Morningstar:1999rf} has been performed 
with anisotropic lattices. The results of \cite{glueb:teper98} and
\cite{Vaccarino:1999ku} at a finite lattice spacing
have been expressed in units of $r_{0}$ in ref. \cite{reviews:tampere}.
In ref. \cite{glueb:teper98} further results are reported for which 
the continuum extrapolation has not been performed.
Results for $m_{0^{++}}r_{0}$ at finite
lattice spacing obtained with the FP (fixed point) action are also present 
\cite{Niedermayer:2000yx}. The error that we give for the average should
not taken too serious, since it is not clear, to which extend the error of 
the individual results is of systematic or statistic nature.
In ref. \cite{MorningstarPeardon2} the authors plot their results 
for $m_{0^{++}} r_0$ obtained
from anisotropic lattices with an improved action, as discussed above. 
They given no final result for the continuum limit. However, from 
the plot one reads off $m_{0^{++}} r_0 \approx 4.0$ with a quite small error;
incompatible with the average of the literature given above by several standard
deviations.

Using the continuum limit relation \cite{SN03}
\begin{equation}
T_c r_0= 0.7498(50)
\end{equation}
the average of the results from the literature can be converted to
\begin{equation}\label{mct0cont}
m_{0^{++}}/T_c|_{a=0} =5.73(9) \;\;.
\end{equation}
We made no attempt to extract a continuum result from our data, since it 
is quite clear from fig. \ref{mass_scaling} that corrections beyond 
$a^2$ are large. 
At $a\simeq 0.11 {\rm fm}$
we do  observe a moderate reduction of the lattice
artefacts by using $\beta_a<0$ with respect to the usual Wilson action
($\beta_a=0$). For $\beta_a=0$, the deviation from the continuum result of
eq.~(\ref{mct0cont}) amounts to $\sim 18\%$,
while for $\beta_a=-2,-4$ it
slightly decreases to  $\sim 12\%$.\\
At $a\simeq 0.17 {\rm fm}$ one observes
discretization errors of $\sim 40\%$ for the Wilson action, while for the
mixed action they amount to  $\sim 25\%$ for $\beta_a=-2$ and $\sim 20\%$ for
$\beta_a=-4$.

\section{Summary and conclusions}
We investigated the SU(3) lattice gauge model
with a pure gauge action that contains plaquette terms in the fundamental
and adjoint representation. In particular, we studied negative 
values of the adjoint coupling $\beta_a$.
This choice is motivated by the presence of a first order phase transition
line in the $(\beta_f,\beta_a)$ plane: one expects that moving towards
negative $\beta_a$ the presence of the endpoint at
$(\beta_f,\beta_a)=(4.00(7),2.06(8))$ becomes less important and hence that
scaling and/or topological properties could be improved with respect to the
Wilson action.
These features would be highly desirable in view of upcoming simulations 
for fermionic actions with exact lattice chiral symmetry and in general for
unquenched computations.

In section 4, we computed the critical coupling
$\beta_{f,c}$ of the finite temperature deconfinement transition at
$1/(aT_c)=2,3,4,6$ and $\beta_a=0,-2,-4$ fixed.
Since this measurement turned to be our most accurate, we have used
$T_c$ to set the scale. We find that lines of constant physics, 
as predicted by one-loop perturbation theory, completely fail to describe 
our numerical results for negative $\beta_a$.

We then calculated the static quark potential from the Polyakov loop
correlation funtion. To this end, we have implemented a variant 
of the algorithm recently proposed by L\"uscher and Weisz. In addition 
to the factorisation in temporal direction, we have employed a 
factorisation in the spacial directions.
This algorithm allowed us to compute the Polyakov loop correlation funtion
up to $r \approx 1.5 {\rm fm}$. Due to these large  distances we were able to
extract the string tension $\sigma$ with little systematic errors (section 5).

Studying the scaling behaviour of the quantity $T_c/\sqrt{\sigma}$ for the
different $\beta_a$ at our disposal, we did not observe a significant
improvement at negative adjoint couplings in comparison to the Wilson case
$\beta_a=0$. The values obtained with negative $\beta_a$ are a little
closer to the continuum limit. 

In section 6 we computed the $0^{++}$ glueball mass for several lattice
spacings and for $\beta_a=0,-2,-4$. Also for this computation we made use of
the factorisation method to reduce the variance of 
the correlation function \cite{meyer}. Although the efficiency
here is not as spectacular as for the Polyakov loop correlation function, 
we were able
to obtain a good statistical accuracy up to distances roughly twice the
ones reached with the standard method.

It turns out that the mass $m_{0^{++}}$ of the lightest glueball is more 
sensitive to the variation of $\beta_a$. This had to be expected, since
at the endpoint of the line of first order phase transitions
the mass in lattice units is zero \cite{Heller}.
Therefore, in particular for $m_{0^{++}}$
large lattice artefacts should show up in the neighbourhood of the endpoint.
We investigated the scaling behaviour of the dimensionless quantity
$m_{0^{++}}/T_c$. Here indeed, we observed a significant reduction of the
lattice artefacts for negative $\beta_a$. At $a\simeq 0.17\, {\rm fm}$,
the lattice artefacts for $\beta_a=0$ are $~40\%$, while for $\beta_a=-4$ they
decrease to $~20\%$.

In view of future dynamical QCD simulations, it would be interesting 
to study the effect on  dislocations  and to investigate spectrum
of the Wilson-Dirac matrix obtained with the mixed action at negative 
$\beta_a$.
Finally one should note that the hybrid-Monte-Carlo (HMC) algorithm 
can be easily implemented for the mixed fundamental-adjoint action.

\section{Acknowledgements}
M.H. thanks PPARC for support under the grant
PPA/G/O/2002/00468. S.N. is supported by TMR, 
EC-Contract No. HPRNCT-2002-00311 (EURIDICE).
Part of the study was conducted,
while the authors have been members of the NIC and Theory group at DESY Zeuthen.
We thank DESY and NIC/DESY for computational resources.
We are grateful to R. Sommer for discussions in the initial
phase of the project.
We thank M. L\"uscher for providing us unpublished data for the
Polyakov loop correlation function. We are grateful to 
F. Gliozzi and O. Ogievetsky for advice on group-theory.


\appendix
\section{Appendix: Is the transfer matrix positive?}
In ref. \cite{transfermatrix} the transfer matrix for lattice 
QCD with the Wilson (fundamental) gauge action is constructed. 
It is straightforward to generalise this construction to
the mixed fundamental/adjoint plaquette action. 

In ref. \cite{transfermatrix} it is shown that the transfer matrix 
for the Wilson action is strictly positive if and only if 
\begin{equation}
\int \mbox{d} U \int \mbox{d} U'
 f^*(U)  \;
\exp\left(\frac{\beta_f}{2 N} [\mbox{Tr}(U^{-1} U') +
\mbox{Tr}(U^{-1} U')^{\dag}]
\right) f(U') > 0 \;
\end{equation}
for all square integrable, nonvanishing functions $f$ on the gauge group 
SU(3).

This can be generalized as 
\begin{equation}
\label{basic}
\int \mbox{d} U \int \mbox{d} U'  
 f^*(U)  \;
\exp\left(\frac{\beta_f}{2 N} [\mbox{Tr} V + 
\mbox{Tr} V^{\dag}]
+ \frac{\beta_a}{N^2} \mbox{Tr} V \; 
\mbox{Tr} V^{\dag} \right) \;
 f(U') > 0
\end{equation}
where $V=U^{-1} U'$.

Following ref. \cite{transfermatrix},
the integration kernel of eq.~(\ref{basic}) 
can be expanded in a Fourier series on the group
\begin{equation}
\label{character}
 \exp\left(\frac{\beta_f}{2 N} [\mbox{Tr} V +
\mbox{Tr} V^{\dag}]
+ \frac{\beta_a}{N^2} \mbox{Tr} V \;
\mbox{Tr} V^{\dag} \right) \;
= \; \sum_{\nu} c_{\nu} \chi^{(\nu)}(V) \;,
\end{equation}
where the sum runs over the set of all irreducible representations 
of SU(3) and $\chi^{(\nu)}(V)$ is the character of the representation $\nu$.
In order that eq.~(\ref{basic}) holds, 
it is necessary and sufficient  that all coefficients $c_{\nu} > 0$ 
are positive. For $\beta_f > 0$ and $\beta_a = 0$ it is proven  
\cite{transfermatrix} that this is indeed the case: 
\begin{equation}
\exp\left(\frac{\beta_f}{2 N} [\mbox{Tr} V +
\mbox{Tr} V^{\dag}] \right)
\; = \; 
\sum_{n,m=0}^{\infty} a_{nm} (\mbox{Tr} V)^n (\mbox{Tr} V^\dag)^m \;\;,
\end{equation}
where $a_{nm}>0$.
Here, $(\mbox{Tr} V)^n (\mbox{Tr} V^\dag)^m$ is the trace of the tensor product
representation of SU(3) composed of $n$ quark and $m$ antiquark representations.
Reducing out the tensor product, one gets
\begin{equation}
 (\mbox{Tr} V)^n (\mbox{Tr} V^\dag)^m \; = \;
\sum_{\nu} c_{\nu}(n,m) \chi^{(\nu)}(V) \;\;,
\end{equation} 
where $c_{\nu}(n,m) > 0$. Since all irreducible representations can be obtained
by reducing out tensor products of quark representations, $c_{\nu} > 0$ for 
all $\nu$. 
It is trivial 
to extend this prove to   $\beta_f > 0$ and $\beta_a \ge 0$.

However for $\beta_a < 0$, as we consider here, the situation becomes 
more complicated. In the expansion
\begin{equation}
\label{expansion}
\exp\left(\frac{\beta_f}{2 N} [\mbox{Tr} V +
\mbox{Tr} V^{\dag}]
+ \frac{\beta_a}{N^2} \mbox{Tr} V \;
\mbox{Tr} V^{\dag} \right) \;
= \sum_{n,m=0}^{\infty} a_{nm} (\mbox{Tr} V)^n (\mbox{Tr} V^{\dag})^m
\end{equation}
it is no longer guaranteed that $a_{nm} > 0$ for all choices of $n,m$. 
This can be most easily seen for 
\begin{equation}
a_{n,1} = \frac{1}{(n+1)!} (n+1) \left[\frac{\beta_f}{2 N} \right]^{n+1} 
        + \frac{1}{n!} n 
           \left[\frac{\beta_f}{2 N} \right]^{n-1} \frac{\beta_a}{N^2} \;\;.
\end{equation}
It follows that $a_{n,1}$ is positive for $(\beta_f/2)^2/n + \beta_a >0$.
I.e. as
$n$ increases, the lower limit on $\beta_a$ goes to zero.
However, it remains quite unclear how the $a_{n,m}$ will add up in the 
coefficients $c_{\nu}$ and whether $c_{\nu}>0$.
We were not able to clarify this question rigorously. 

To get some idea, we evaluated the coefficients $c_{\nu}$ for the 
pairs of $\beta_f, \beta_a$ studied in this paper,
for representations $\nu$ up to the dimension 15.

To this end we evaluated the integrals
\begin{equation}
 c_{\nu} = \int \mbox{d} V \chi_{\nu}^*(V) 
\exp\left(\frac{\beta_f}{2 N} [\mbox{Tr} V +
\mbox{Tr} V^{\dag}]
+ \frac{\beta_a}{N^2} \mbox{Tr} V \;
\mbox{Tr} V^{\dag} \right) \;
\end{equation}
numerically. It turned out that $c_{\nu}>0$ for all coefficients
that we computed, except for the coefficient of one 15 dimensional 
representation at $\beta_f=7.8477$ and $\beta_a=-4$.

We applied a second numerical approach to check the positivity of 
the integration kernel. 
Assume that we  perform the integration eq.~(\ref{basic}) 
with a Monte Carlo method. I.e. we evaluate $f(U)$
for $m$ SU(3) matrices that have been selected randomly.
Choosing the same SU(3) matrices 
for both integrations, the integration kernel becomes a real symmetric 
$m \times m$ matrix. In this study, we have used $m \le 10000$.

To check the positivity of this matrix, we evaluated its smallest eigenvalue.
It turned out that for $\beta_a=-2$ and $\beta_a=-4$ and all values 
of $\beta_f$ that we have studied here, negative eigenvalues are found.
However, in particular for $\beta_a=-2$, the absolute value of the smallest
eigenvalue is by  several orders of magnitude smaller than the largest 
eigenvalue. Notice that no quantitative sign of a violation of 
positivity has been observed in the decay of the Wilson loop correlation
functions, while for the improved 
actions studied in ref. \cite{SN03} such violations
were clearly visible.
It remains an open question, whether there is a finite range of negative 
$\beta_a$, where the transfermatrix is strictly positive.


\end{document}